\documentclass[11pt]{elsart1p}

\usepackage{natbib}
\usepackage{epsfig}
\usepackage{amssymb}
\journal{New Astronomy}

\newcommand{\bl}[1]{\mbox{\boldmath$ #1 $}}
\newcommand{\vnp}{\bl{v}_{\rm n}}
\newcommand{\vip}{\bl{v}_{\rm i}}
\newcommand{\cs}{c_{\rm s}}
\newcommand{\csq}{c_{\rm s}^2}
\newcommand{\ceffsq}{\tilde{C}_{\rm eff}^2}
\newcommand{\ceffsqi}{\tilde{C}_{\rm eff,0}^2}

\newcommand{\beq}{\begin{equation}}
\newcommand{\eeq}{\end{equation}}
\newcommand{\barray}{\begin{eqnarray}}
\newcommand{\earray}{\end{eqnarray}}

\newcommand{\Msun}{M_{\odot}}

\newcommand{\bfig}{\begin{figure}}
\newcommand{\efig}{\end{figure}}
\newcommand{\FT}{{\cal F}}
\newcommand{\FTinv}{{\cal F}^{-1}}

\newcommand{\cmc}{~{\rm cm}^{-3}}
\newcommand{\cms}{~{\rm cm}^{-2}}
\newcommand{\Alf}{Alfv\'en\ }
\newcommand{\Alfvc}{Alfv\'enic\ }
\newcommand{\Alfv}{Alfv\'enic}

\newcommand{\kms}{~\rm km~s^{-1}}

\newcommand{\pc}{~\rm pc}

\newcommand{\K}{~\rm K}
\newcommand{\muG}{~\mu{\rm G}}
\newcommand{\yr}{~{\rm yr}}

\newcommand{\AU}{~{\rm AU}}
\newcommand{\mH}{m_{\rm H}}
\newcommand{\Pext}{P_{\rm ext}}
\newcommand{\Pexttil}{\tilde{P}_{\rm ext}}
\newcommand{\mui}{\mu_0}

\newcommand{\tni}{\tau_{\rm ni}}
\newcommand{\tnii}{\tau_{\rm ni,0}}
\newcommand{\tniitil}{\tilde{\tau}_{\rm ni,0}}
\newcommand{\trun}{t_{\rm run}}
\newcommand{\kion}{k_{\rm i}}
\newcommand{\rhon}{\rho_{\rm n}}
\newcommand{\rhoni}{\rho_{\rm n,0}}
\newcommand{\nn}{n_{\rm n}}

\newcommand{\nni}{n_{\rm n,0}}
\newcommand{\nion}{n_{\rm i}}
\newcommand{\sign}{\sigma_{\rm n}}

\newcommand{\signi}{\sigma_{\rm n,0}}
\newcommand{\signmax}{\sigma_{\rm n,max}}
\newcommand{\Beq}{B_{z,\rm eq}}

\newcommand{\Bref}{B_{\rm ref}}

\newcommand{\vmax}{v_{\rm max}}
\newcommand{\vms}{V_{\rm MS,0}}
\newcommand{\vasqi}{\tilde{V}_{\rm A,0}^2}

\newcommand{\lammax}{\lambda_{\rm T,m}}
\newcommand{\Htwo}{{\rm H}_{2}}
\newcommand{\sigin}{\langle \sigma w \rangle_{\rm{i\Htwo}}}
\newcommand{\ekin}{E_{\rm kin}}
\newcommand{\ekini}{E_{\rm kin,0}}
\newcommand{\Nni}{N_{\rm n,0}}
\newcommand{\xion}{x_{\rm i}}
\newcommand{\mn}{m_{\rm n}}
\newcommand{\mi}{m_{\rm i}}

\begin{document}

\begin{frontmatter}

\title{Magnetically-Regulated Fragmentation Induced by Nonlinear Flows
and Ambipolar Diffusion}

\author[label1]{Shantanu Basu\corauthref{cor}}, 
\corauth[cor]{Corresponding author.}
\ead{basu@astro.uwo.ca}
\author[label2]{Glenn E. Ciolek},
\ead{cioleg@rpi.edu}
\author[label1]{Wolf Dapp},
and \author[label1]{James Wurster}

\address[label1]{Department of Physics and Astronomy, University of
Western Ontario, London, Ontario N6A~3K7, Canada}
\address[label2]{New York Center for Astrobiology and Department of Physics, Applied Physics, and Astronomy,
Rensselaer Polytechnic Institute, 110 W. 8th Street, Troy, NY 12180,
USA}

\begin{abstract}
We present a parameter study of simulations of fragmentation
regulated by gravity, magnetic fields, ambipolar diffusion, and 
nonlinear flows. The thin-sheet approximation is employed 
with periodic lateral boundary conditions, and the 
nonlinear flow field (``turbulence'') is allowed to freely decay.
In agreement with previous results in the literature,
our results show that the onset of runaway collapse (formation of
the first star) in subcritical clouds is significantly accelerated 
by nonlinear flows in which a large-scale wave mode dominates the
power spectrum. 
In addition, we find that a power spectrum with equal energy on all
scales also accelerates collapse, but by a lesser amount.
For a highly super-\Alfvc initial velocity field with
most power on the largest scales, the
collapse occurs promptly during the initial compression 
wave. However, for trans-\Alfvc perturbations, a 
subcritical magnetic field causes a rebound from the initial compression,
and the system undergoes several oscillations before runaway 
collapse occurs. Models that undergo prompt collapse
have highly supersonic infall motions at the core boundaries.
Cores in magnetically subcritical models with trans-\Alfvc 
initial perturbations also pick up significant systematic 
speeds by inheriting motions associated with magnetically-driven
oscillations. Core mass distributions are much broader than in
models with small-amplitude initial perturbations, although the
disturbed structure of cores that form due to nonlinear flows 
does not guarantee subsequent monolithic collapse. 
Our simulations also demonstrate that significant power can
(if present initially) be maintained with negligible 
dissipation in large-scale compressive modes of a magnetic thin sheet, 
in the limit of perfect flux freezing.


\end{abstract}


\end{frontmatter}

\section{Introduction}

Magnetic fields and supersonic turbulence are two mechanisms
that are commonly invoked for the regulation of
star formation in our Galaxy to the observationally
estimated rate of $\sim 3-5\, \Msun$ yr$^{-1}$ \citep[see][]{mck89}.
This is at least one hundred 
times less than the rate implied by the gravitational fragmentation timescale
of the molecular gas in the Galaxy calculated from its mean density.
Put another way, the global Galactic star formation efficiency (SFE) is about 
1\% of molecular gas per free-fall time. Interestingly, the same SFE
also applies to nearby individual star-forming regions such as the
Taurus molecular cloud \citep{gol08}.
The relatively low Galactic SFE is one fundamental
constraint on the global properties of star formation. The
existence of a broad-tailed core mass function (CMF) that is 
a lognormal with a possible power-law high-mass tail, is another. In fact, the
observed form of the CMF \citep[e.g.][]{mot98} is similar to that of 
the stellar initial mass function, the IMF.
Other important star formation constraints specifically applying to 
cores include the generally subsonic relative infall motions
\citep{taf98,wil99,lee01,cas02}, the low
(subsonic or transonic) systematic core speeds
\citep{and07,kir07}, and the somewhat non-circular
projected shapes \citep{mye91,jon01}. The relatively low speeds are an important property
since the cores are embedded in molecular clouds whose overall 
internal random motions are highly supersonic.

Since most stars form in clusters or loose groups, 
it seems clear that some sort of 
fragmentation process is at work in those regions.
There are several qualitatively distinct 
modes of fragmentation to consider. The simplest process is gravitational 
fragmentation, which can be
divided into cases with mass-to-flux ratios that are
supercritical (gravity-dominated)
and subcritical (ambipolar-diffusion-driven). 
An alternate and distinct star formation mechanism is 
turbulent fragmentation, dominated by nonlinear flows,
which can also occur in clouds with supercritical and subcritical 
mass-to-flux ratios.
The limit of highly supercritical clouds also corresponds
to super-\Alfvc turbulence in the case that turbulent and gravitational
energies have comparable magnitude. This limit has been advocated by
\citet{pad02}. However, we favor the transcritical and trans-\Alfvc
cases on theoretical and observational grounds, as discussed
in Section 4.
For a more complete understanding of star formation, a study of the 
interplay of magnetic fields, turbulence, and ambipolar diffusion is 
therefore of great importance.
In this paper, we carry out an extensive parameter survey of 
magnetic field strengths and initial nonlinear perturbations, 
facilitated by our use of the thin-sheet approximation, and discuss the 
consequences of our results. Such broad parameter studies remain out of
reach for fully three-dimensional non-ideal MHD simulations \citep{kud08,nak08}.

In a previous paper (\citealt{bas09}, hereafter BCW; see also \citealt{bas04}), 
we studied the nonlinear
evolution of gravitational fragmentation initiated by small-amplitude 
perturbations, including the effects of magnetic fields and
ambipolar diffusion. An extensive parameter study was performed,
encompassing the supercritical,
transcritical, and subcritical cases, and also accounting for varying
levels of cloud ionization and external pressure. Some main findings
of that paper were that fragment spacings in the nonlinear phase agree
with the predictions of linear theory \citep{cio06}, and that the time to 
runaway collapse
from small-amplitude white-noise initial perturbations is up to ten times
the growth time calculated from linear theory. 
\citet{cio06} showed that transcritical
gravitational fragmentation can lead to significantly larger size and mass scales than
either the supercritical or subcritical limits.
BCW found that CMFs for regions with 
a single uniform initial mass-to-flux ratio are sharply peaked, but that 
the sum of results from simulations with a variety of initial mass-to-flux
ratios near the critical value can produce a broad distribution.
This represents a way to get a broad CMF, of the type observed, 
without the need for nonlinear forcing. 
Importantly, only a {\it narrow} initial distribution of initial mass-to-flux
ratio is needed to create a relatively {\it broad} CMF.
Additionally, BCW showed that different ambient mass-to-flux ratios in 
different regions lead to observationally distinguishable values of
infall motions, core shapes, and magnetic field line curvature.
\citet{kud07} have recently performed three-dimensional
simulations of gravitational fragmentation with magnetic fields and
ambipolar diffusion, and verified some of the main findings of the thin-sheet
models. Particularly, they also found the dichotomy between subsonic maximum infall
speeds in subcritical clouds and somewhat supersonic speeds in 
supercritical clouds. 

The inclusion of nonlinear (hereafter, ``turbulent'') initial conditions
to fragmentation models including ambipolar diffusion was introduced
by \citet{li04} and \citet{nak05}, employing the thin-sheet approximation. 
They found that the timescale of 
star formation was reduced significantly by the initial motions with
power spectrum $v_k^2 \propto k^{-4}$ \citep{li04}, to become 
$\sim 10^6$ yr for an initially somewhat subcritical cloud rather than
$\sim 10^7$ yr. By continuing to integrate past the collapse of the
first core through the use of an artificially stiff equation of state
for surface densities 10 times greater than the initial value, 
they found that magnetic fields nevertheless prevented most material
from collapsing to form stars. 
\citet{kud08} have verified that the mode
of turbulence-accelerated magnetically-regulated star formation also
occurs in a fully three-dimensional simulation.
While three-dimensional simulations are resource-limited,
and large parameter studies cannot yet be performed, 
\citet{kud08} showed that this mode of star formation proceeds
though an initial phase of enhanced ambipolar diffusion created by the
small length scale generated by the large-scale compression associated with
the initial perturbation. If this is not sufficient to raise the 
maximum mass-to-flux ratio above the critical value, there is a 
rebound to lower densities. However, the highest density regions 
remain well above the initial mean density, and here the ambipolar
diffusion proceeds in a quasistatic manner ($\propto \rhon^{-1/2}$ assuming
force balance between gravity and magnetic forces - see \citealt{mou99})
but at an enhanced rate due to the raised density.


In this paper, we study the effect of large-amplitude nonlinear initial perturbations
on the evolution of a thin sheet whose evolution is regulated by
magnetic fields and ambipolar diffusion. We focus on
the early stages of prestellar core formation and evolution, and
do not integrate past the runaway collapse of the first core.
Therefore, the effects of protostellar feedback through
outflows do not need to be added. These simplifications allow us to 
run a large number of simulations. We perform an
extensive parameter study and also study many realizations of models
with a single set of parameters, since the initial turbulent state is 
inherently random. 
Some important questions that we can address and which have not been 
answered in previous papers are as follows. In which parameter space do
nonlinear velocity fields
lead to prompt collapse, and in which cases can magnetic fields cause a
rebound from the first compressions? 
How sensitively does the time until runaway collapse 
depend upon the values of different parameters? What is the effect of
different power spectra of perturbations? Is there a qualitative 
difference between \Alfvc and super-\Alfvc perturbations? 
How do velocity profiles in the vicinity of cores vary in the 
different scenarios? What are the systematic speeds of cores?

%

Our paper is organized in the following manner. 
The model is described in Section 2, results are given in Section 3,
and a discussion of results, including speculation and implications for
global star formation in a molecular cloud are given in Section 4.
We summarize our results in Section 5.

\section{Physical and Numerical Model}

We employ the magnetic thin-sheet approximation, as laid out in detail
in several previous papers \citep{cio93,cio06,bas09}. 
Physically, we are modeling the dense mid-layer of a molecular cloud, and
ignoring the more rarefied envelope of the cloud.
We assume isothermality at all times.
The basic equations
governing the evolution of a model cloud (conservation of 
mass and momentum, Maxwell's equations, etc.) are integrated 
along the vertical axis from $z=-Z(x,y)$ to $z=+Z(x,y)$. In doing so, 
a ``one-zone approximation'' is used, in which all quantities
are taken to be independent of height within the sheet. 
The volume density of neutrals $\rhon$ is calculated from the vertical pressure balance
equation
\beq
\rhon \csq = \frac{\pi}{2}G \sign^2 + \Pext + \frac{B_x^2+B_y^2}{8\pi},
\eeq 
where $\cs$ is the isothermal sound speed, $\sign(x,y) = \int_{-Z}^{+Z}\rhon(x,y)~dz$ is 
the column density of neutrals, $\Pext$ is the external pressure on the sheet,
and $B_x$ and $B_y$ represent values of magnetic field components 
at the top surface of the sheet, $z=+Z$.

We solve normalized versions of the magnetic thin-sheet
equations. The unit of velocity is taken
to be $\cs$, the column density unit is $\signi$, 
and the unit of acceleration is $2 \pi G \signi$, equal to the 
magnitude of vertical acceleration above the sheet. Therefore, 
the time unit is $t_0 = \cs/2\pi G \signi$, and the length unit is 
$L_0= \csq/2 \pi G \signi$. From this system we can also construct 
a unit of magnetic field strength, $B_0 = 2 \pi G^{1/2} \signi$. 
The unit of mass is $M_0 = \cs^4/(4\pi^2G^2\,\signi)$.
Here, $\signi$ is the uniform 
neutral column density of the background state.
Typical values of our units are given in the Appendix.
With these normalizations, the equations used to determine the
evolution of a model cloud are
\barray
\label{cont}
 \frac{{\partial \sign }}{{\partial t}} & = & - \nabla_p  \cdot \left( \sign \, \vnp \right), \\
\label{mom}
 \frac{\partial}{\partial t}(\sign \vnp)  & = & - \nabla_p \cdot (\sign \vnp \vnp) +  \bl{F}_{\rm T}
+ \bl{F}_{\rm M} + \sign \bl{g}_p, \\
\label{induct}
 \frac{{\partial \Beq }}{{\partial t}} & = & - \nabla_p  \cdot \left( \Beq \, \vip \right) , \\
\label{Ftherm}
\bl{F}_{\rm T} & = & - \ceffsq \nabla_p \sign                   ,\\
\label{Fmag}
\bl{F}_{\rm M} & = & \Beq \, ( \bl{B}_p - Z\, \nabla_p \Beq ) + {\cal O}(\nabla_p Z), \\
\label{vieq}
\vip & = & \vnp + \frac{\tniitil}{\sign}\left(\frac{\rhoni}{\rhon}\right)^{\kion} \bl{F}_{\rm M} , \\  
\label{ceffeq}
\ceffsq & = & \sign^2 \frac{(3 \Pexttil + \sign^2)}{(\Pexttil + \sign^2)^2} , \\
\label{rhon}
\rhon & = & \frac{1}{4} \left( \sign^2 + \Pexttil + \bl{B}_p^2 \right),\\
\label{Zeq}
Z & = & \frac{\sign}{2 \rhon}, \\                                
\bl{g}_p & = & -\nabla_p \psi  ,\\
\label{gravpot}
\psi & = & \FTinv \left[ - \FT(\sign)/k_z \right]     ,\\
\bl{B}_p & = & -\nabla_p \Psi  ,\\
\label{magpot}
\Psi & = & \FTinv \left[ \FT(\Beq - \Bref)/k_z \right]  \, .
\earray
In the above equations, 
$\bl{B}_p(x,y) = B_x(x,y)\hat{\bl{x}} + B_y(x,y)\hat{\bl{y}}$ is the 
planar magnetic field at the top surface of the sheet, 
$\vnp(x,y) = v_x(x,y)\hat{\bl{x}} + v_y(x,y)\hat{\bl{y}}$ is the 
velocity of the neutrals in the plane, and 
$\vip(x,y) = v_{{\rm i},x}(x,y)\hat{\bl{x}} + v_{{\rm i},y}(x,y)\hat{\bl{y}}$ 
is the corresponding velocity of the ions.
The operator $\nabla_p = \hat{\bl{x}} \, \partial/\partial x + 
\hat{\bl{y}} \, \partial/\partial y$ is the gradient in the planar
directions within the sheet.
The quantities $\psi(x,y)$ and $\Psi(x,y)$ are the 
scalar gravitational and magnetic potentials in the plane
of the sheet, derived in the limit that the sheet is infinitesimally
thin. The vertical wavenumber $k_z = (k_x^2+k_y^2)^{1/2}$ is 
a function of wavenumbers $k_x$ and $k_y$ in the plane of the sheet,
and the operators $\FT$ and $\FTinv$ represent the forward
and inverse Fourier transforms, respectively, which we calculate 
numerically using an FFT technique. Terms of order 
${\cal O}(\nabla_p Z)$ in $\bl{F}_{\rm M}$, the magnetic force per
unit area, are not 
written down for the sake of brevity, but are included in the numerical
code; their exact form is given in Sections 2.2 and 2.3 of \citet{cio06}. 
All terms proportional to $\nabla_p Z$ are generally very small.

The Eq. (\ref{vieq}) above makes use of relations for the
neutral-ion collision time $\tni$ and the ion density $\nion$,
as described in BCW:
\barray
\label{tni}
\tni & = & 1.4 \,\frac{\mi + m_{{}_{\Htwo}}}{\mi} \frac{1}{\nion \sigin}\;, \\
\label{nion}
\nion &= & {\cal K} \nn^{\kion}.
\earray
Furthermore, Eq. (\ref{vieq}) uses
the normalized initial mass density (in units of $\signi/L_0$)
$\rhoni = \frac{1}{4}(1+\Pexttil)$, where $\Pexttil$ is defined below. 

Our basic equations contain three dimensionless free parameters:
$\mui \equiv 2 \pi G^{1/2} \signi/\Bref$ is the dimensionless
(in units of the critical value for gravitational collapse) 
mass-to-flux ratio of the reference state; 
$\tniitil \equiv \tnii/t_0$ is the dimensionless neutral-ion collision
time of the reference state, and expresses the effect of ambipolar
diffusion;
$\Pexttil \equiv 2 \Pext/\pi G\signi^{2}$ is the ratio of the external
pressure acting on the sheet to the vertical self-gravitational stress
of the reference state.

Each numerical model is carried out
within a square computational domain spanning the
region $-L/2 \leq x \leq L/2$ and $-L/2 \leq y \leq L/2$. 
Periodic boundary conditions are imposed in the $x$ and $y$ directions.
We choose a domain size $L=16\pi\,L_0$ for all models presented 
in this paper, so that
it is four times wider than the wavelength
of maximum growth rate for an isothermal nonmagnetic and unpressured
sheet, $\lammax = 4\pi\,L_0$ 
(see \citealt{cio06}; BCW).
The computational domain has $N$ zones in each direction, with
$N=128$ unless stated otherwise. Some results utilizing greater $N$
are presented in Section \ref{s:tdiss}.

Gradients in our simulation box are approximated using 
second-order accurate central differencing.
Advection of mass and magnetic flux is prescribed by using the monotonic
upwind scheme of \citet{van77}. These forms of spatial discretization
convert the system of partial differential equations to a system of
ordinary differential equations (ODE's), with one ODE for each
physical variable at each cell. Time-integration of this large coupled
system of ODE's is performed by using an Adams-Bashforth-Moulton
predictor-corrector subroutine \citep{sha94}. Numerical solution of
Fourier transforms and inverse transforms, necessary to calculate the
gravitational and magnetic potentials $\psi$ and $\Psi$ at each time
step (see Eqs. [\ref{gravpot}] and [\ref{magpot}]), is done by using
fast Fourier transform techniques.
Further details of our numerical technique are given by BCW.

The initial conditions of our model include a ``turbulent'' velocity field
added to our background state of uniform column density $\signi$
and vertical magnetic field strength $\Bref$. The velocity field is
generated in Fourier space using the usual practice adopted by e.g.
\citet{sto98} for three-dimensional models and \citet{li04} for thin-sheet
models. For a physical grid consisting of $N$ zones in each $(x,y)$
direction, the wavenumbers $k_x$ and $k_y$ in Fourier space consist of the  
discrete values $k_j = 2\pi j/L$, where $j=-N/2,...,N/2$. 
For each pair of $k_x$ and $k_y$, we assign a 
Fourier velocity amplitude chosen from a Gaussian distribution and 
scaled by the power spectrum $v_k^2 \propto k^{n}$, where
$k = (k_x^2+k_y^2)^{1/2}$. The phase is chosen from a uniform random
distribution in the interval $[0,2\pi]$. The resulting Fourier velocity
field is then transformed back into physical space. The distributions
of each of $v_x$ and $v_y$ are chosen independently in this manner, 
and each is rescaled so that the rms amplitude of the field is equal to $v_a$.
For $n=-4$, the resulting velocity field has most of its power in a large-scale
flow component. We have experimented
with various values of $n$ and find that the results are  
generally similar as long as $n < 0$. 
Distinct results are found in the case of flat spectrum perturbations ($n=0$),
which we present for comparison.
Finally, we note that velocity fields generated in the
above manner are compressive. Hence, we have also studied one
model with $n=-4$ but the Fourier space amplitudes chosen so that
$v_x$ and $v_y$ satisfy  $\nabla_p  \cdot \vnp =0$.

Therefore, our turbulent initial conditions introduce the additional 
dimensionless free parameters $v_a/\cs$ and $n$, while our simulation box 
introduces the parameters $L/\lammax$ and grid size $N$.

\section{Results}

\begin{table}
\begin{center}\caption{Models and Parameters}\end{center}
\begin{tabular}{crrrcccr}
\hline \hline
Model &\hspace{2em}$\mu_0$ &\hspace{2em}$\tniitil$ &\hspace{2em}$\Pexttil$ &\hspace{2em}Spectrum &\hspace{1em}$v_{\rm a}/\cs$ &\hspace{2em}$\vms/\cs$ &\hspace{2em}$\trun/t_0$ \\
\hline 
1 	&0.5	 &0.0	 &0.1	 &$k^{-4}$	& 2.0 &2.9   &$>5000$\\
2 	&0.5	 &0.2	 &0.1	 &$k^{-4}$	&4.0  &2.9   &0.8\\
3 	&0.5	 &0.2	 &0.1	 &$k^{-4}$	&3.0  &2.9  &30\\
4 	&0.5	 &0.2	 &0.1	 &$k^{-4}$	&2.0  &2.9  &31\\
5 	&0.5	 &0.2	 &0.1	 &$k^{-4}$	&1.0  &2.9  &50\\
6 	&0.5	 &0.2	 &0.1	 &$k^{-4}$	&0.5  &2.9  &58\\
7 	&0.5	 &0.2	 &10.0 	 &$k^{-4}$  	&2.0  &1.0 &8.5\\
8 	&0.5	 &0.2	 &0.1	 &$k^{-4}$ (div0)&2.0 &2.9 &23\\
9 	&0.5	 &0.2	 &0.1	 &$k^0$		&2.0  &2.9 &56\\
10 	&0.5	 &0.1 	 &0.1 	 &$k^{-4}$	&2.0  &2.9 &92\\
11 	&0.5	 &0.4	 &0.1 	 &$k^{-4}$	&2.0  &2.9 &8.1\\
12 	&0.5	 &1.0	 &0.1	 &$k^{-4}$	&2.0  &2.9 &1.8\\
13	&1.0	 &0.2	 &0.1	 &$k^{-4}$	&2.0  &1.7 &1.8\\
14	&1.0	 &0.2	 &0.1	 &$k^{-4}$	&1.0  &1.7 &4.3\\
15	&1.0	 &0.2	 &0.1	 &$k^{-4}$	&0.5  &1.7 &12\\
16	&1.0	 &0.2	 &0.1	 &$k^0$		&2.0  &1.7 &33\\
17	&2.0	 &0.2	 &0.1	 &$k^{-4}$	&2.0  &1.2 &1.3\\
\hline
\end{tabular}
\end{table}

Table 1 contains a summary of models in our parameter study. 
For each model, we list the values of the free parameters
$\mui, \tniitil, \Pexttil$, the
form of the turbulent power spectrum, the turbulent velocity amplitude
$v_a$, the magnetosound speed $\vms$ in the initial state of the model, 
and the calculated duration of the simulation, $\trun$.
The initial magnetosound speed is calculated from the other parameters
of the model, and its relation to $v_a$ can act as a useful diagnostic.
Following \citet{cio06} we use
\beq
\vms = \left( \vasqi + \ceffsqi \right)^{1/2}\,\cs,
\eeq
where $\vasqi = 2 \mui^{-2}/(1+\Pexttil)$ is the 
square of the normalized (to $\cs$) initial \Alf speed, and
$\ceffsqi = (1+3 \Pexttil)/(1 + \Pexttil)^2$ is the square of a
normalized initial effective sound speed that includes the effect of external
pressure, and follows from Eq. (\ref{ceffeq}).
Model 7 has $\ceffsqi < 1$ because for $\Pexttil=10$, the large external
pressure contributes significant opposition to the restoring force of 
internal pressure in the initial state \citep[see discussion in][]{cio06}.

Our simulations are terminated as soon as $\signmax \geq 10\, \signi$, corresponding
to runaway gravitational collapse of the first core. 
For models with $\Pexttil = 0.1$, this also corresponds to a volume
density enhancement by a factor $\approx 100$. We have verified with
high resolution runs to greater values of $\signmax/\signi$ that the collapse
does indeed continue. Collapsing regions are also invariably 
gravity-dominated, having locally 
supercritical mass-to-flux ratios and net accelerations that point toward
the density peak. This includes cases of
prompt collapse, i.e. when collapse occurs in localized regions 
due to strong shocks associated with the turbulent flow field, in a time
$\trun$ less than $2\,t_0$.
The values of $\trun$ do
vary somewhat from one realization of the initial state to another,
and in many cases represent an average value from many simulations. 
We have run each model at least five separate times, and some
have been run significantly many more times, as described below.
Model 1, which evolves under flux-frozen conditions ($\tniitil=0$), cannot be 
terminated in the usual manner.
Since the initial mass-to-flux ratio is also
subcritical for this model ($\mui=0.5$), gravitational runaway collapse
is not possible unless there is significant numerical magnetic
diffusion. That simulation ran past $t= 5000\,t_0$ without runaway
collapse or any notable artificial flux dissipation, thus providing
an excellent verification of the accuracy of our code.  
While some models undergo many oscillations before
eventual runaway collapse of density peaks, others undergo prompt collapse. 
Although the models that undergo
prompt collapse may be considered to be artificially forced into collapse
by large-scale flows in the initial conditions, we nevertheless 
present them here as interesting limiting cases.
Models 4, 13, and 17 constitute a special set of models with our standard
neutral-ion coupling parameter $\tniitil=0.2$, external pressure
$\Pexttil=0.1$,
turbulent velocity amplitude $v_a=2.0\,\cs$, but varying initial mass-to-flux
ratio parameter $\mui=0.5,1.0,2.0$, respectively. 
Model 4 is run 15 times, and models 13 and 17 are run 25 times 
(with unique random realizations
of the initial velocity field), in order to compile statistics on the
core mass distributions using the techniques described by BCW.
Model 4 can in some sense be considered our ``standard'' model since we are
most interested in the acceleration of collapse in subcritical clouds
due to nonlinear supersonic velocity perturbations.
Model 8 is similar to model 4 but has the divergence-free initial velocity
field. Model 3 is on the brink of either prompt collapse or
a longer-term evolution leading to runaway collapse, and can sometimes
go into prompt collapse (see Section 4). For this reason, we ran the 
model over 15 times 
in order to yield a $\trun = 30\,t_0$ that is a characteristic value for all
but a few runs that do go into prompt collapse.

\subsection{Global Properties}

\bfig
\centering
\begin{tabular}{cc}
\epsfig{file=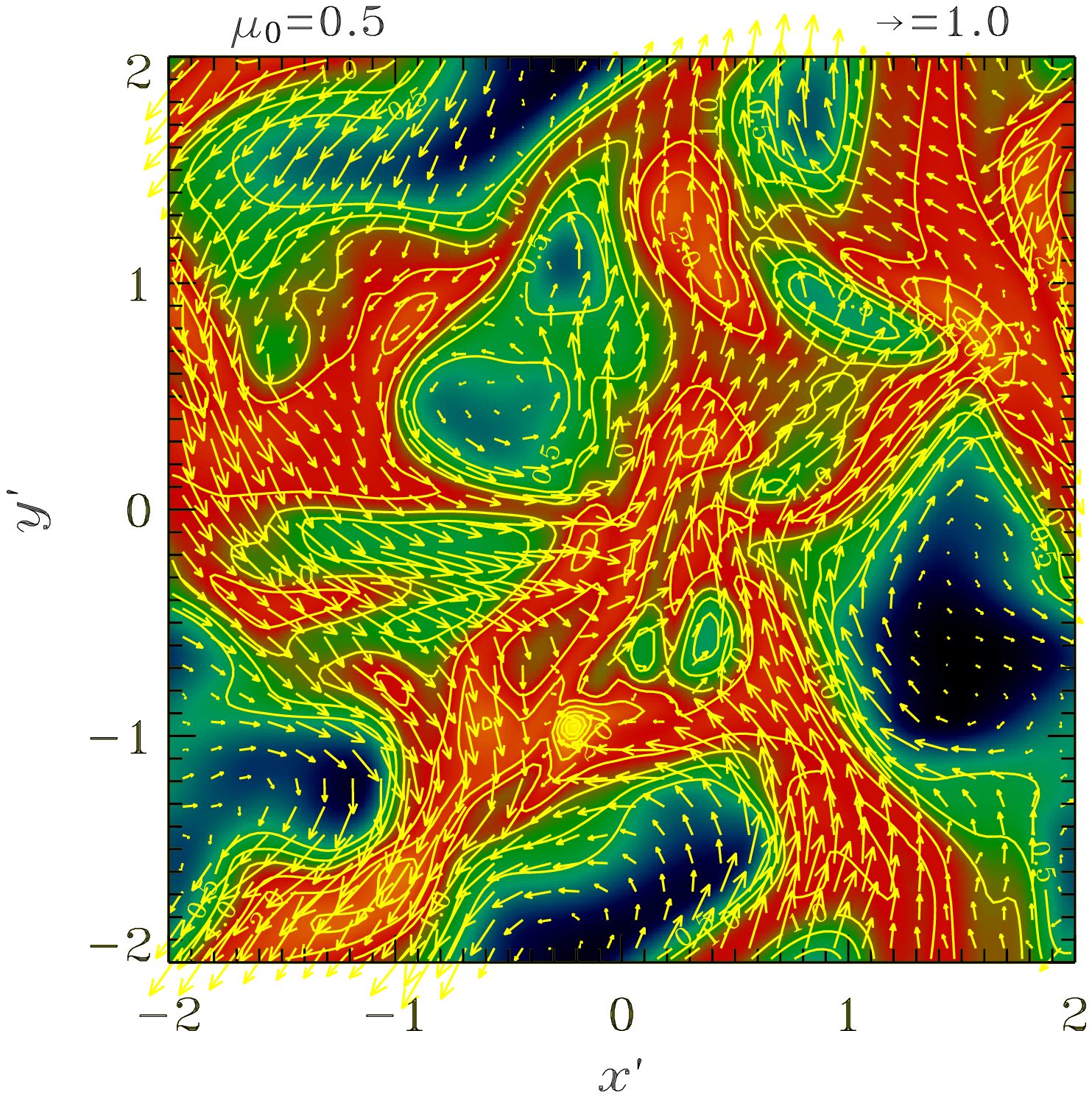,width=0.5\linewidth,clip=} &
\epsfig{file=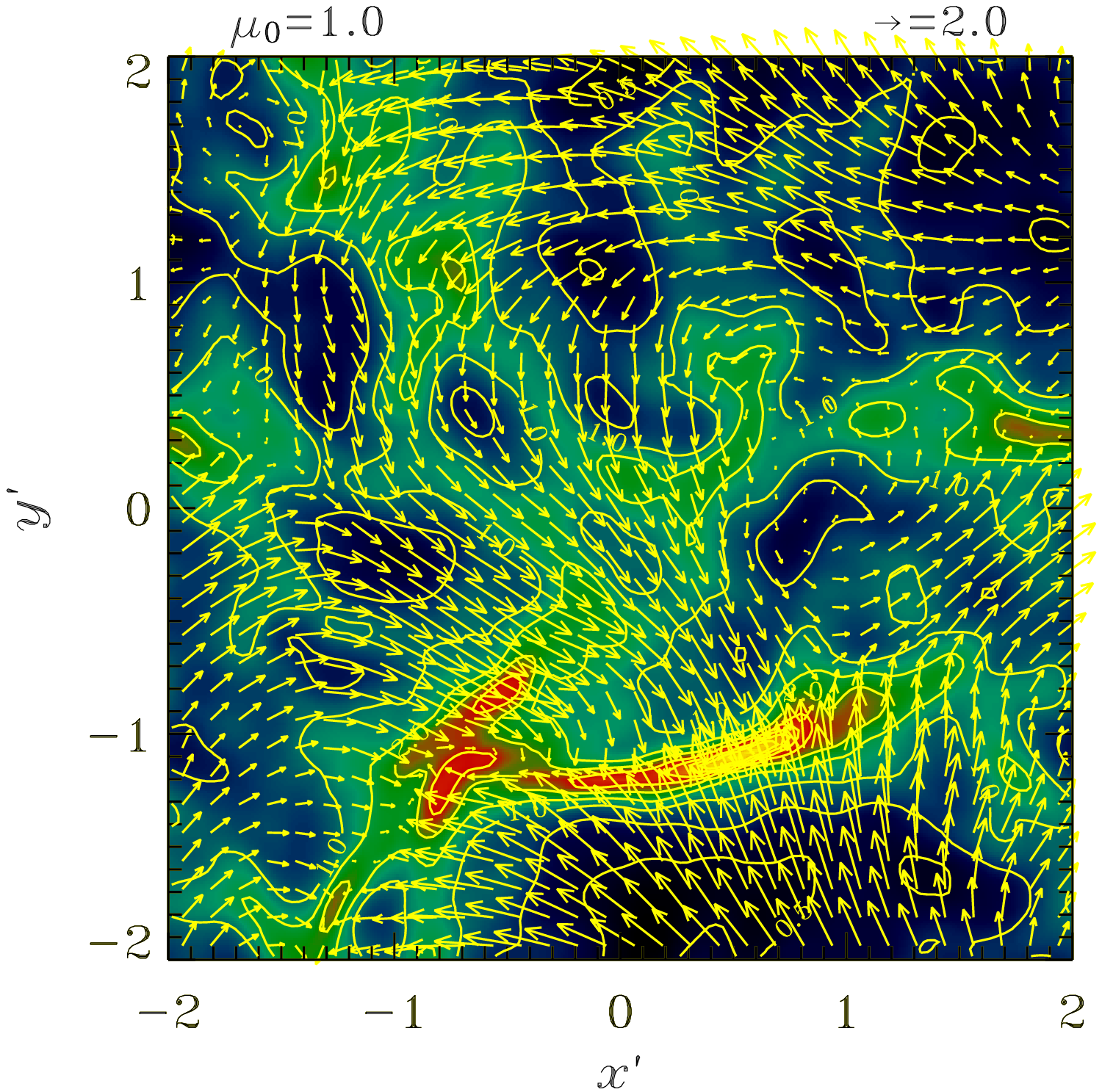,width=0.5\linewidth,clip=} \\
\epsfig{file=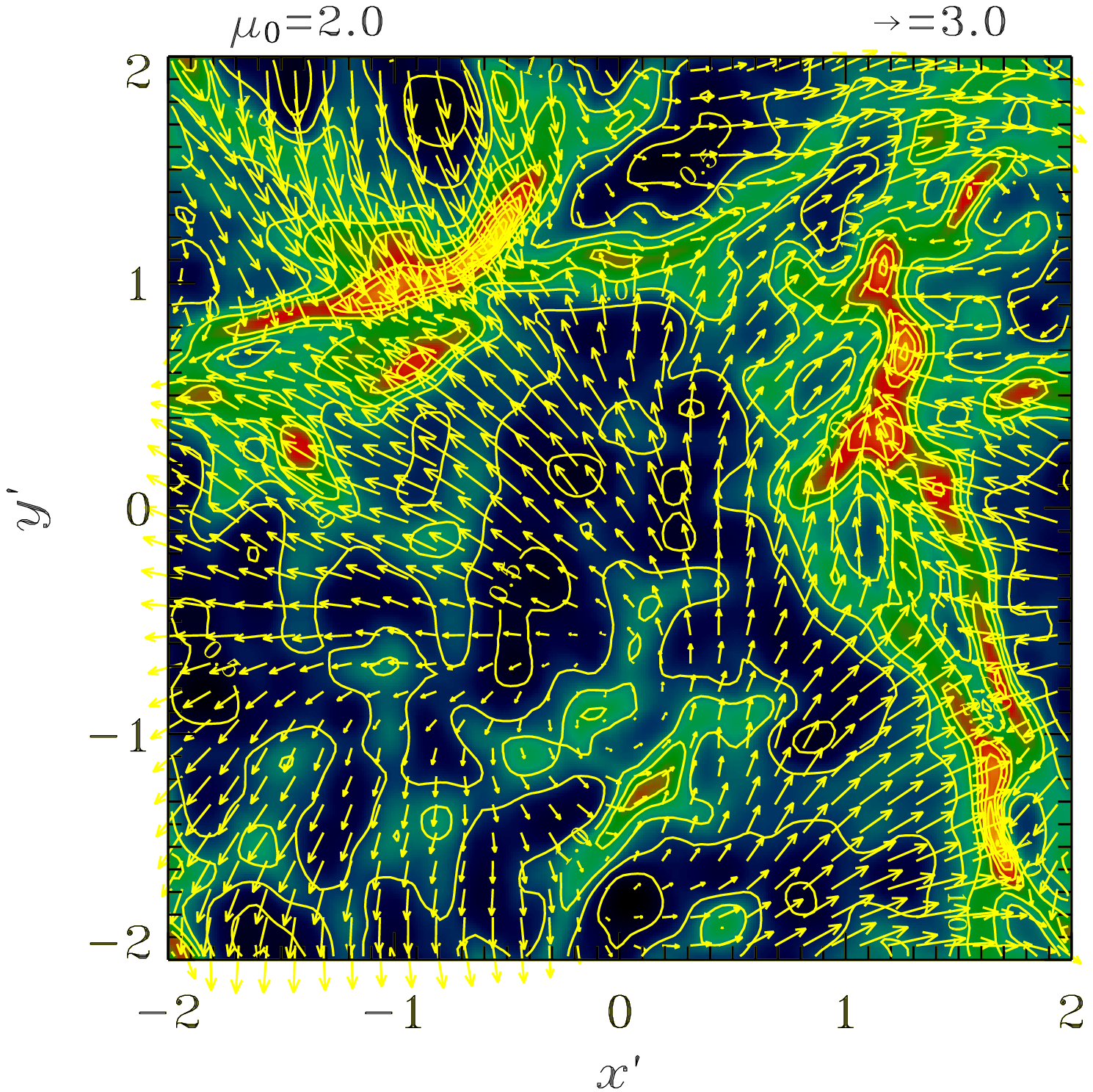,width=0.5\linewidth,clip=} &
\end{tabular}
\caption{
Image and contours of column density $\sign(x,y)/\signi$,
and velocity vectors of neutrals, for three different models at the time
that $\signmax/\signi = 10$. All models have $\tniitil=0.2$,
$\Pexttil=0.1$, and $v_a=2.0\,\cs$. Top left: model 4 ($\mui=0.5$). 
Top right: model 13 ($\mui=1.0$). 
Bottom left: model 17 ($\mui=2.0$). 
The color table
is applied to the logarithm of the column density and the contour lines
represent values of $\sign/\signi$ spaced in
multiplicative increments of $2^{1/2}$, having the values
[0.7,1.0,1.4,2,2.8,4.0,...].
The horizontal or vertical distance between the footpoints of velocity
vectors corresponds to a speed $1.0 \, \cs$ for the $\mui=0.5$ model,
$2.0 \, \cs$ for the $\mui=1.0$ model, and $3.0 \, \cs$ for the $\mui=2.0$ model.
We use the normalized spatial coordinates
$x' = x/\lammax$ and $y' = y/\lammax$, where $\lammax=4\pi\,L_0$ is the wavelength of
maximum growth rate from linear perturbation theory, in the 
nonmagnetic limit with $\Pext=0$.
}
\label{densimgs}
\efig

Fig.~\ref{densimgs} shows images of column density overlaid with 
column density
contours and velocity vectors, for realizations of models 4 (top left), 
13 (top right), and 17 (bottom left). Each snapshot is at the end of the
simulation, when $\signmax/\signi = 10$, but occurs at a different time
$\trun$, as indicated in Table 1. 
The maximum speeds in the simulation region are quite different at the
end of the three simulations even though all three start with perturbations
characterized by $v_a = 2.0\,\cs$. Therefore, the 
velocity vector plots each have a different normalization, with the 
horizontal or vertical distance between footpoints of vectors corresponding to
$1.0\,\cs$, $2.0\,\cs$, and $3.0\,\cs$ for the three models with $\mui=0.5,1.0$,
and $2.0$ respectively. To understand the difference in maximum speeds, it is
important to understand the different course of evolution in each model.

The model 4, with $\mui=0.5$, has a strong enough magnetic field that the
initial compression driven by the large-scale flow of the nonlinear velocity
field does not lead to prompt collapse in any region. The magnetic
field causes a rebound after the initial compression. The densest regions 
never reexpand fully to the initial background density, and instead undergo
oscillations in density until continuing ambipolar diffusion 
leads to the creation of regions of supercritical mass-to-flux ratio. These
regions then undergo runaway collapse. For model 4, this occurs at 
a representative $t= 31\,t_0$, meaning
that there is sufficient time for the initial velocity field to have damped
significantly, since we do not replenish turbulent energy in these simulations.
This is why the velocity amplitude is much smaller at the end of the simulation
than in the other two runs. However, the maximum value is still supersonic
($3.2\,\cs$), and
there are strong systematic flow fields in the simulation. In contrast, 
when starting with small-amplitude initial perturbations (BCW), 
the maximum infall speeds are subsonic. In that case, runaway collapse also 
occurred much later, at $t=204\,t_0$.
The color table and column density contours for model 4, when compared to
those for models 13 and 17, reveal that the gas is not as compressed and
filamentary as in those cases, due to the rebound from the initial extreme
compressions.  

The images, contours, and velocity vectors for models 13 ($\mui=1.0$)
and 17 ($\mui=2.0$) reveal that the initial strong compression leads to
immediate runaway collapse within highly compressed filaments. The
velocity field has highly ordered supersonic compressive infall motions.
The runaway collapse is occurring at times $t=1.8\,t_0$ and $t=1.3\,t_0$,
respectively, essentially as soon as the initial flow creates a large-scale
compression.  Although kinetic energy is efficiently dissipated behind the
shock fronts, there is hardly enough time for a large reduction of
the global kinetic energy.
Therefore, the maximum speeds at the end of the simulation 
($6.0\,\cs$ for model 13 and $7.2\,\cs$ for model 17) are quite
similar to the initial maximum values.

\bfig
\centering
\begin{tabular}{cc}
\epsfig{file=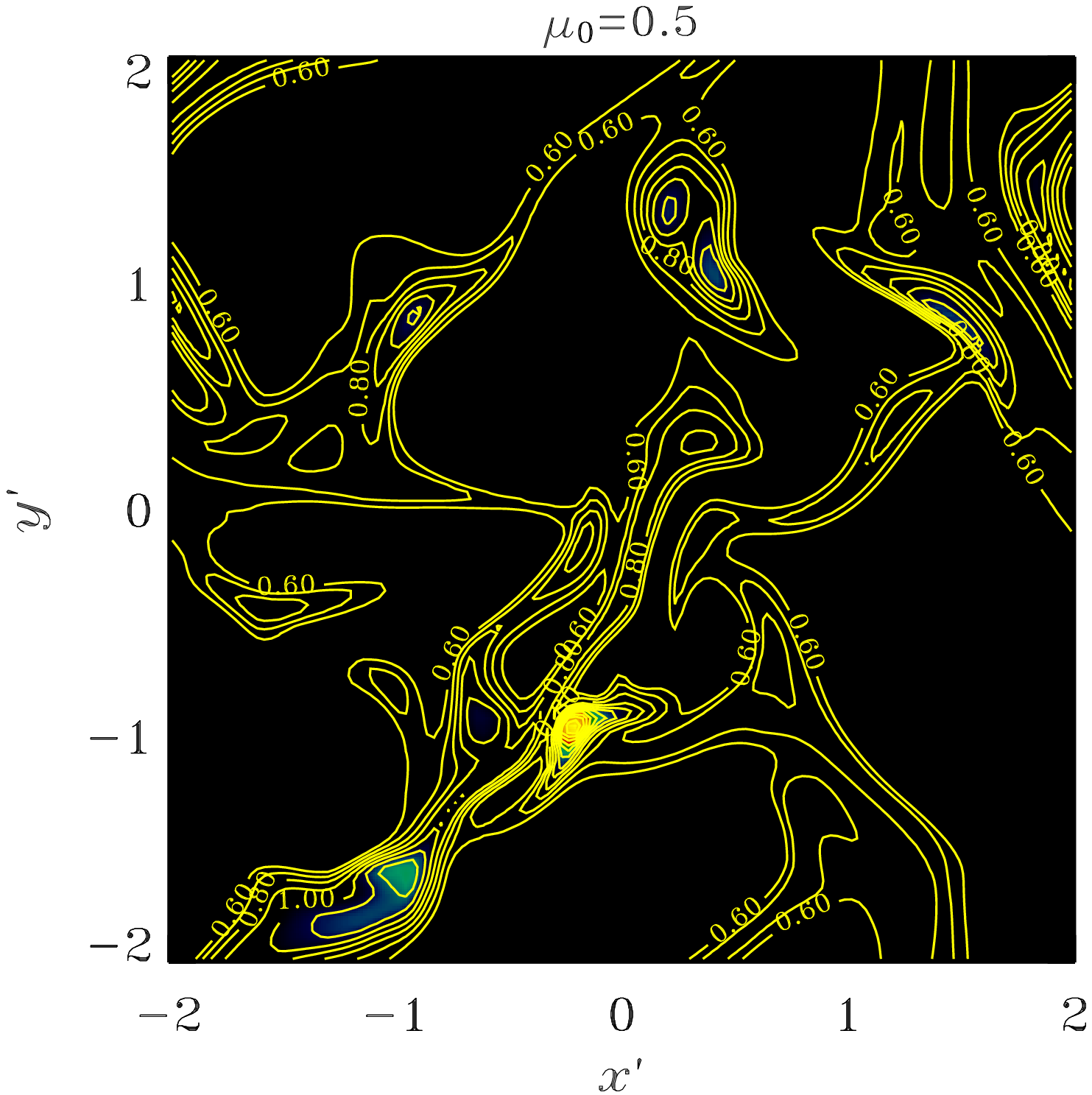,width=0.5\linewidth,clip=} &
\epsfig{file=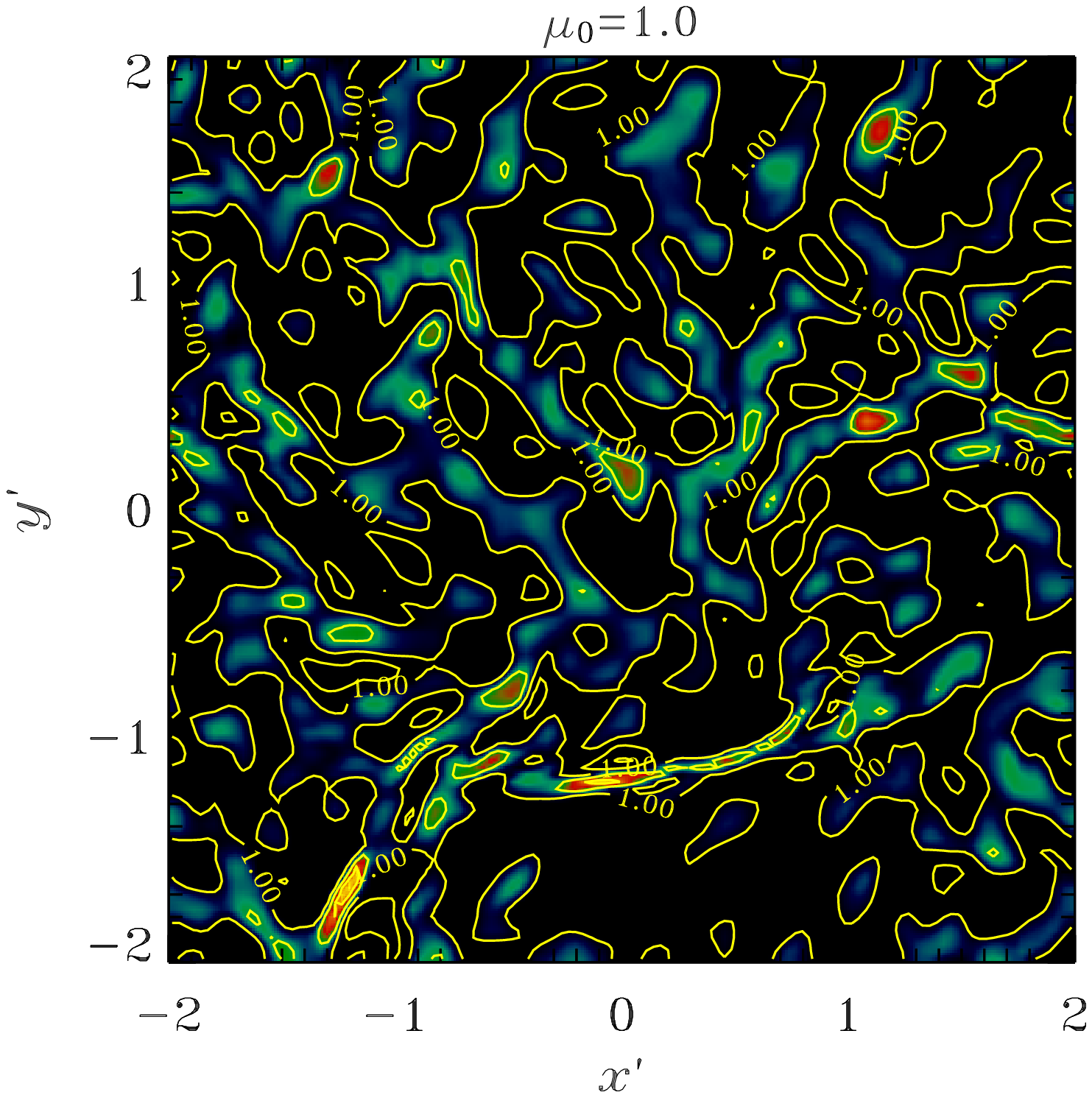,width=0.5\linewidth,clip=} 
\end{tabular}
\caption{
Image and contours of $\mu(x,y)$, the mass-to-flux ratio in units of the
critical value for collapse. Regions with $\mu >1$ are displayed with a color table,
while regions with $\mu <1$ are black. The contour lines are spaced in additive
increments of 0.1. Left: Final snapshot of model 4 ($\mui=0.5$). 
Right: Final snapshot of model 13 ($\mui=1.0$).
}
\label{mtoflx}
\efig

Fig.~\ref{mtoflx} shows images of the mass-to-flux ratio at the end
of the simulations of models 4 and 13. 
The end states have a combination of subcritical and supercritical regions.
Subcritical regions are shown in black, and a color table is applied to the
supercritical regions on both panels. The left panel illustrates that the 
supercritical regions of the initially significantly subcritical ($\mui=0.5$) 
cloud are created within the filamentary regions generated by the large-scale
compressions. In contrast, the initially critical ($\mui=1.0$) model has
widespread supercritical regions generated by the small-scale modes of 
turbulence, as well as the most supercritical regions in the compressed layers.
The former effect of widespread patches of mildly supercritical gas is possible
due to the marginal nature of the critical ($\mui=1.0$) initial state. 
Physically, we might expect
the cloud with $\mui=1.0$ to lead to a cluster of stars soon after the 
runaway collapse of the first cores. On the other hand, the
significantly subcritical cloud with $\mui=0.5$ would show only isolated star
formation in the compressed layers and have to wait a much longer time before
ambipolar diffusion leads to clustered star formation in the remainder of the cloud.

\bfig
\centering
\begin{tabular}{cc}
\epsfig{file=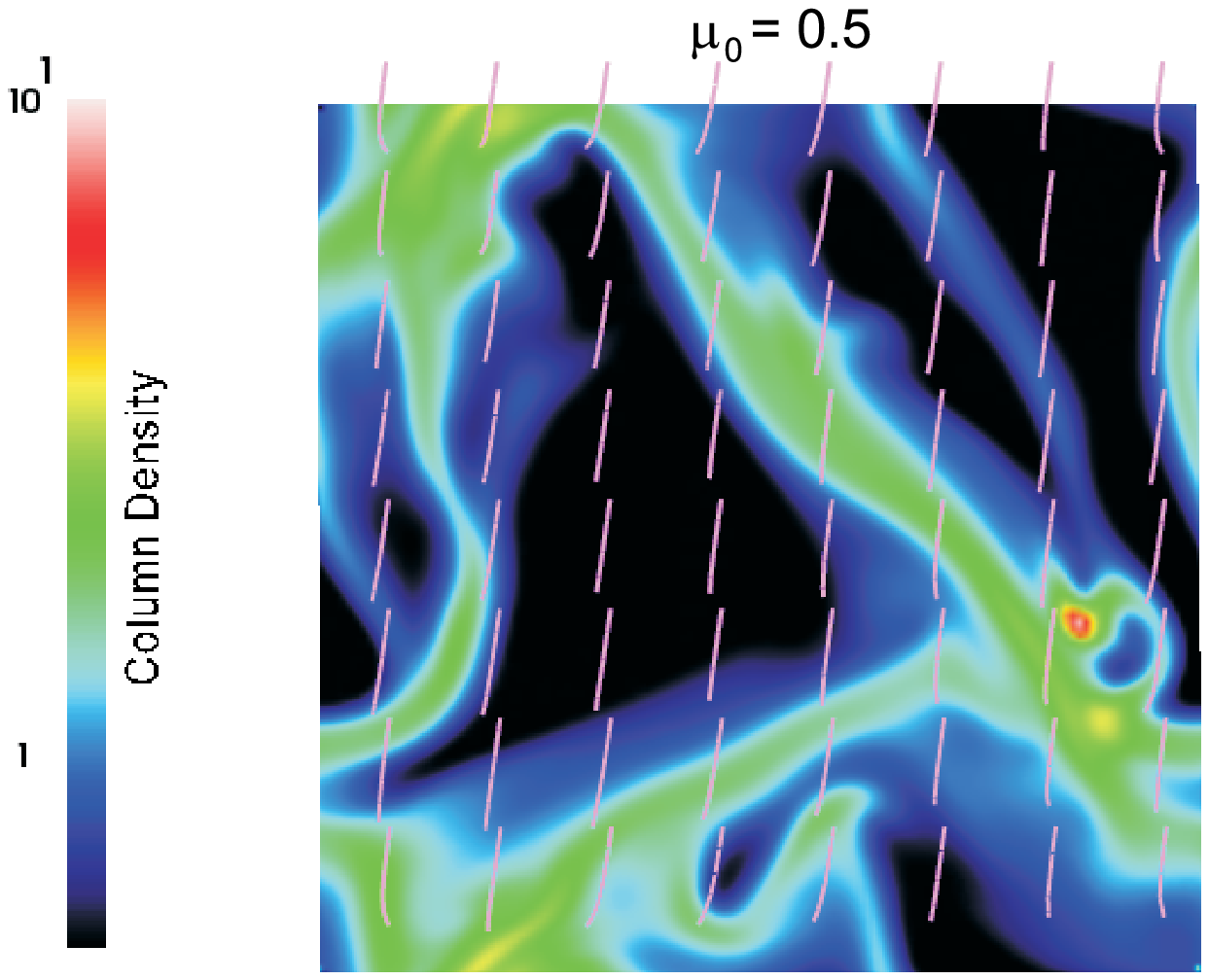,width=0.5\linewidth,clip=} &
\epsfig{file=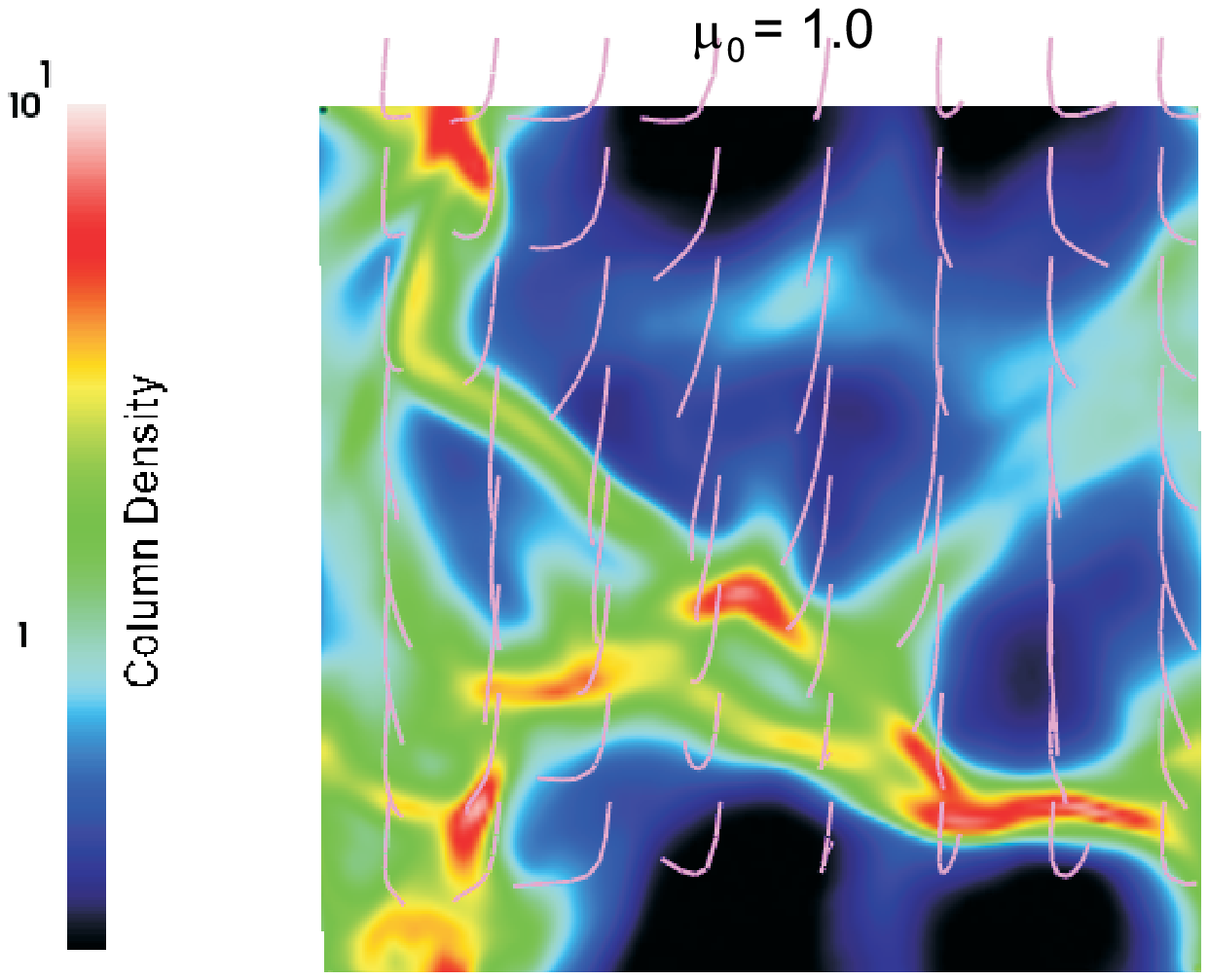,width=0.5\linewidth,clip=} \\
\epsfig{file=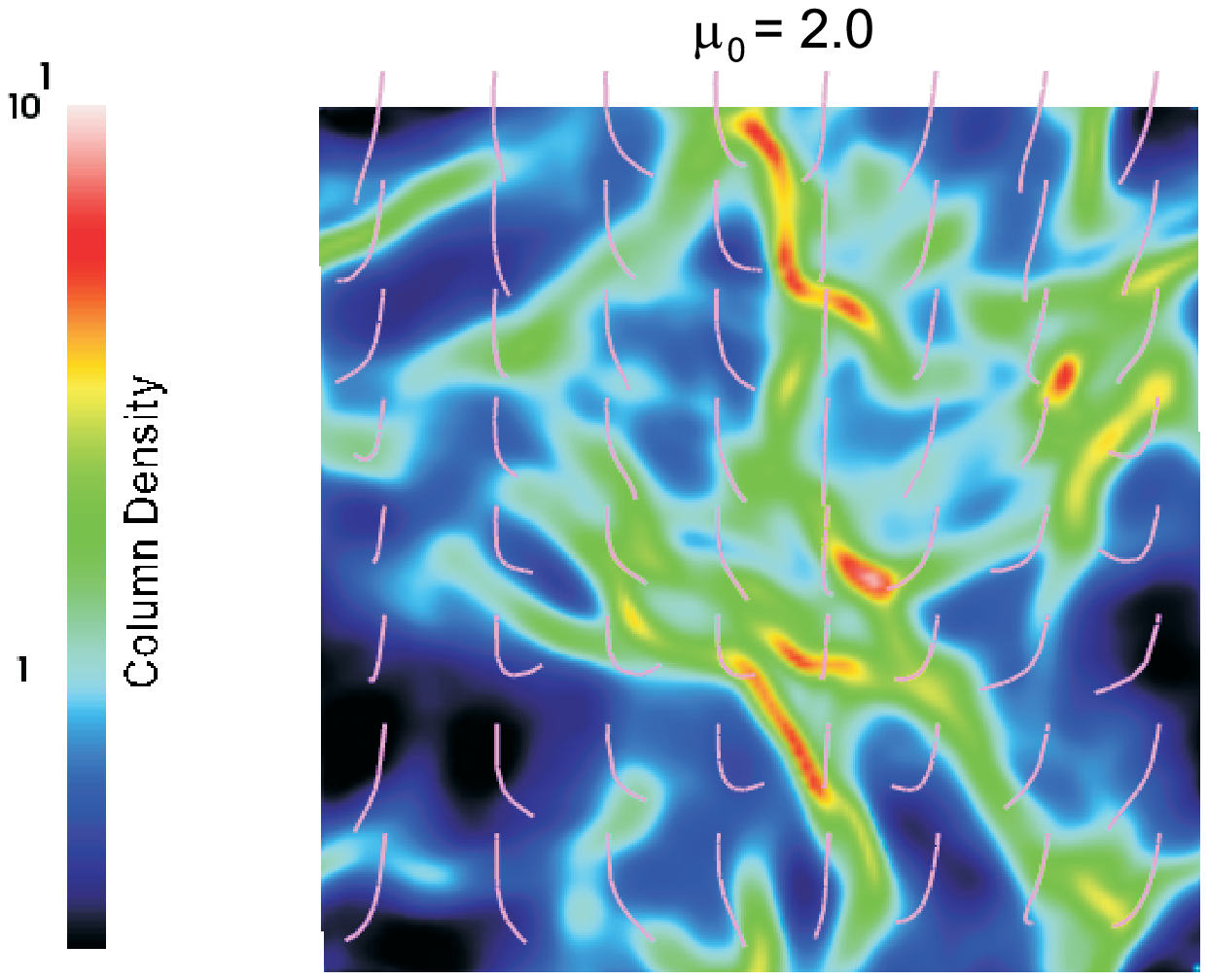,width=0.5\linewidth,clip=} &
\end{tabular}
\caption{
Image of gas column density $\sign(x,y)/\signi$
and superposed magnetic field lines for realizations of 
models 4, 13, and 17, with $\mui=0.5$ (top left), $\mui=1.0$ (top right),
and $\mui=2.0$ (bottom left). All models have initial velocity
amplitude $v_a = 2.0\,\cs$. These are two-dimensional projections
of three-dimensional images containing
a sheet with a column density image and magnetic field lines extending
above the sheet to a distance about half the box width. The image is seen
from a viewing angle of about 10$^{\circ}$ relative to the sheet normal
direction.
Animations of the
evolution of the column density are available online. The field lines
appear in the last frame of the animation.
}
\label{movieimgs}
\efig

Fig.~\ref{movieimgs} shows the end states of models 4, 13, and 17 in 
different realizations (i.e. starting with a different but statistically
equivalent initial velocity field) than shown in Fig.~\ref{densimgs}. 
A color table shows the column density of the final state, and magnetic
field lines above the sheet are also illustrated. These lines are generated
in the manner described in BCW. The image of sheet surface and
field lines above are viewed from an angle of $10^{\circ}$ from the sheet
normal direction. Animations of the evolution of the sheet surface density, with
field lines appearing in the last frame, are available online.
Clearly, the models which suffer prompt collapse ($\mui=1.0$ and $\mui=2.0$)
show the most curvature of field lines since the field is dragged inward
by the strong initial compression wave. In contrast, the cloud with 
$\mui=0.5$ (top left) undergoes a rebound and several oscillations before
runaway collapse can occur. This allows the magnetic field to
straighten out again. The ultimate collapse of the first core is due
to ambipolar drift of neutrals past field lines, so the field is not
significantly distorted by this process. However, a legacy of the 
initial compression is that the mass-to-flux ratio is no longer 
spatially uniform, and significant column density structure exists
even if the magnetic field is not very distorted.

The relative amounts of field line curvature in the cloud and within dense
cores are quantified by $\theta = \tan^{-1} (|B_p|/\Beq)$,
where $|B_p| = (B_x^2+B_y^2)^{1/2}$ is the magnitude of the planar magnetic
field at any location on the sheet-like cloud. Hence, $\theta$ is the angle that
a field line makes with the vertical direction at any location at the top
or bottom surface of the sheet. To quantify the differences in field line
bending from subcritical to transcritical to supercritical clouds,
we note that representative realizations of models 4, 13, and 17, 
with $\mui=(0.5,1.0,2.0)$, have average
values $\theta_{\rm av} = (4.2^{\circ}, 18^{\circ}, 23^{\circ})$, and maximum
values (probing the most evolved core in each simulation)
$\theta_{\rm max} = (25^{\circ}, 65^{\circ}, 65^{\circ})$.
Of these, only the model with $\mui=0.5$ shows similar values of $\theta$
as a corresponding model with initial small-amplitude perturbations. For
models with small-amplitude initial perturbations, BCW found that 
$\mui=(0.5,1.0,2.0)$ yields representative values
$\theta_{\rm av} = (1.7^{\circ}, 8.3^{\circ}, 18^{\circ})$, and 
$\theta_{\rm max} = (20^{\circ}, 30^{\circ}, 46^{\circ})$.

\bfig
\centering
\begin{tabular}{cc}
\epsfig{file=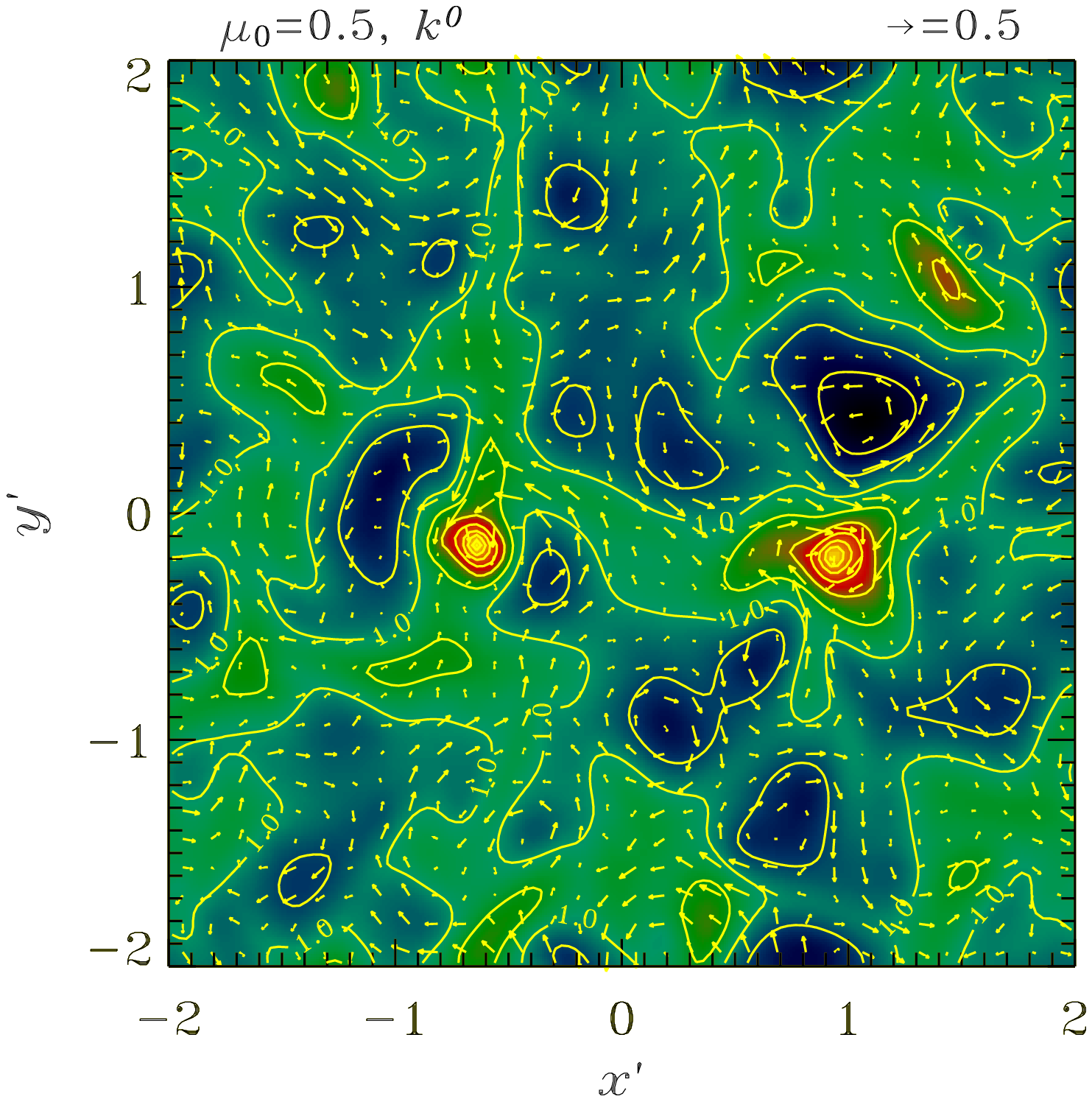,width=0.5\linewidth,clip=} &
\epsfig{file=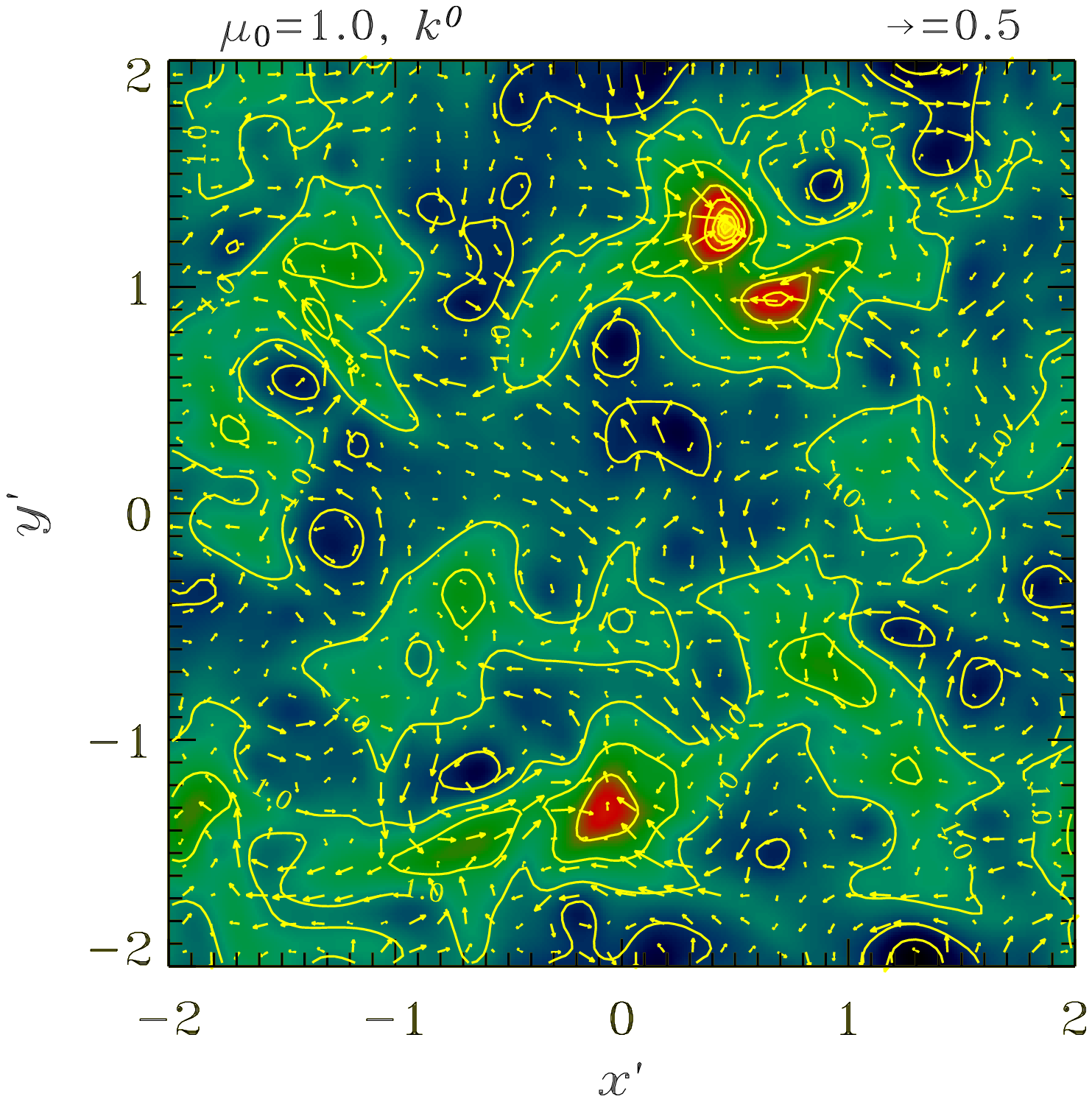,width=0.5\linewidth,clip=} \\  
\epsfig{file=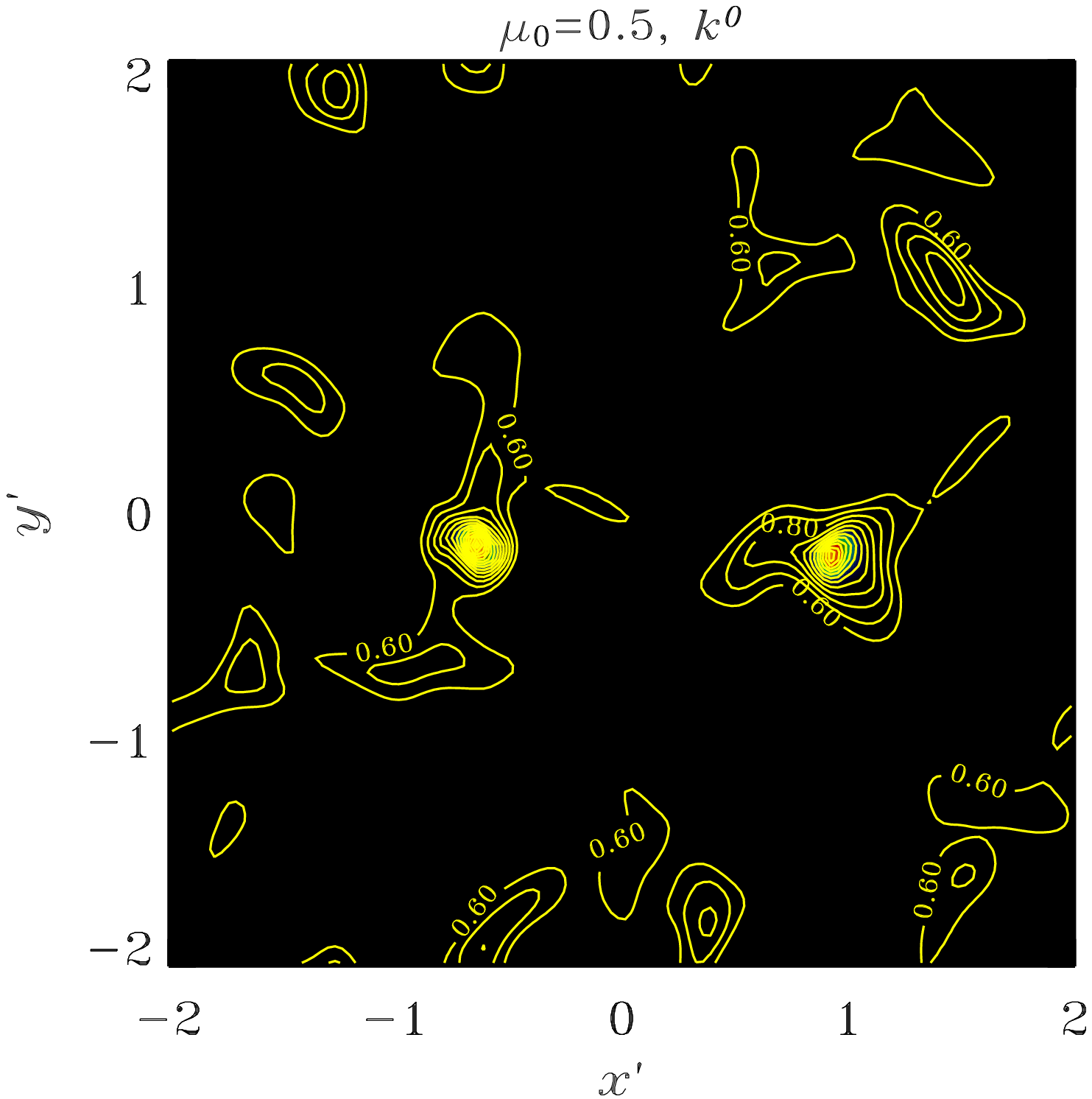,width=0.5\linewidth,clip=} &
\epsfig{file=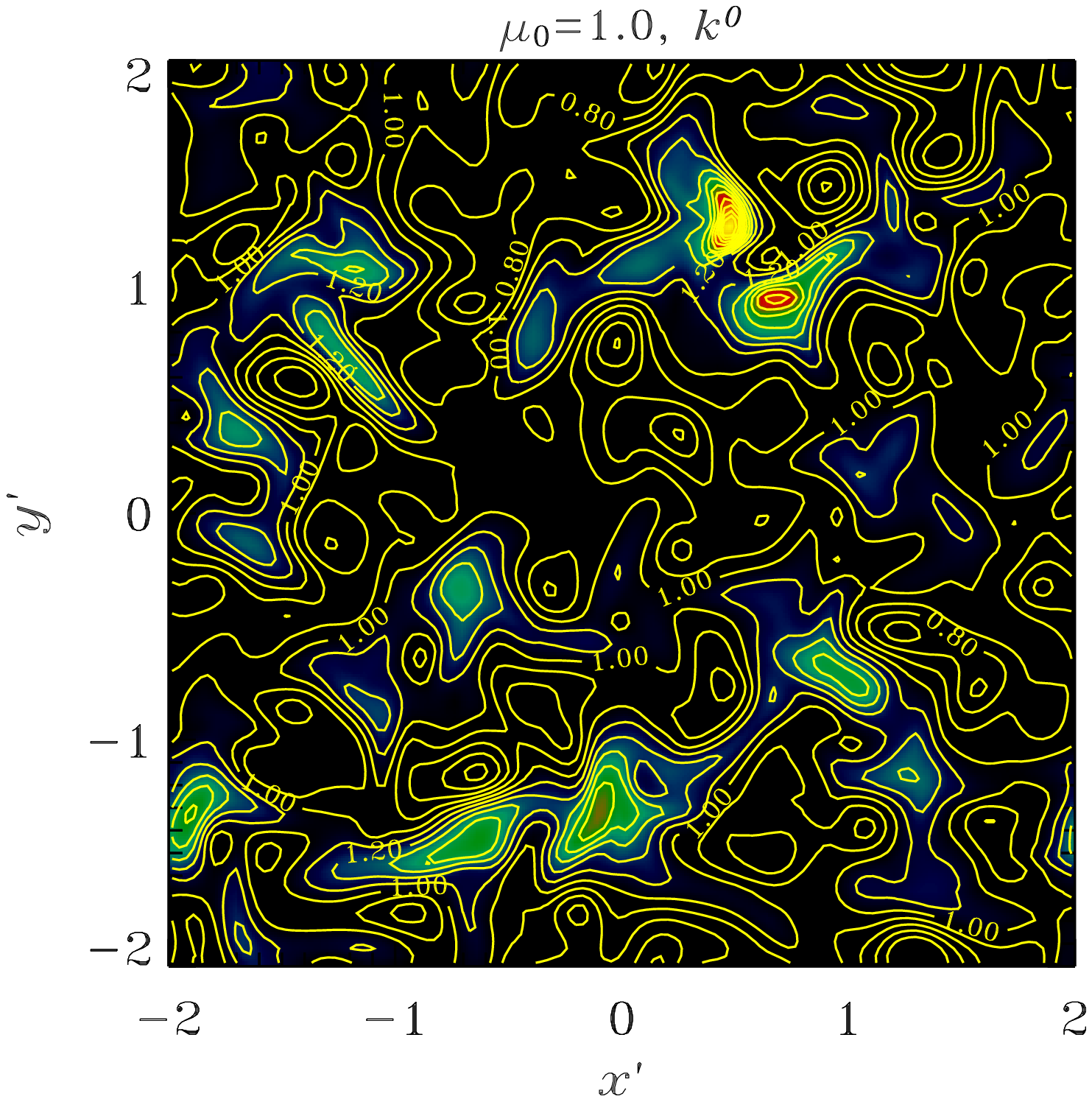,width=0.5\linewidth,clip=} 
\end{tabular}
\caption{
Top: Column density and velocity vectors as in Fig. \ref{densimgs}, but for
models 9 and 16, which have initial nonlinear velocity field with $v_a=2.0\,\cs$ 
and flat power spectrum ($k^0$).
The horizontal or vertical distance between the origins of velocity
vectors corresponds to a speed $0.5 \, \cs$.
Bottom:
Images of mass-to-flux ratio, as in Fig.~\ref{mtoflx} but for 
models 9 and 16.
}
\label{k0densmtfimg}
\efig

The evolution of models with flat spectrum ($v_k^2 \propto k^0$) initial
perturbations is distinct from the cases with negative exponent,
so we present the results from models 9 and 16 in Fig.~\ref{k0densmtfimg}.
In these models, $v_a=2.0\,\cs$ as in model 4, but the large-scale flow does 
not dominate the initial
condition. Therefore, the small-scale modes contribute more 
significantly to enhance
ambipolar diffusion. This enhancement of ambipolar diffusion
due to small-scale irregularities is similar to the mechanism
studied analytically by \citet{fat02} and \citet{zwe02}. 
We study only models with $\mui=0.5$
and $\mui=1.0$ in order to focus on the enhanced ambipolar
diffusion. The values of $\trun$ for these models are
$56\,t_0$ and $33\,t_0$ respectively. These are significantly 
longer time scales
than in the corresponding models with $v_k^2 \propto k^{-4}$.
However, they are much shorter than in models with the same background state
and linear initial perturbations (studied by BCW), in which case 
$\trun$ is $204\,t_0$ and $121\,t_0$ for $\mui=0.5$ and 
$\mui=1.0$, respectively. 

Fig.~\ref{k0densmtfimg} shows the column density and mass-to-flux ratio
at the end of simulations of model 9 and 16.
The input turbulence acts to increase the rate of ambipolar diffusion,
but the turbulence also decays away. By the time of runaway collapse, the
cloud structure and kinematics more closely resembles the case of
small-amplitude initial perturbations than the
case of nonlinear-flow-induced fragmentation (models 4 and 13).
Representative values of the maximum speed $\vmax$ of neutrals 
at the time $t=\trun$ of runaway collapse are
$0.7\,\cs$ for model 9 and $0.8\,\cs$ for model 16. Both values are closer to
the values $0.4\,\cs$ and $0.7\,\cs$ when starting with small-amplitude 
initial perturbations (BCW)
than for the cases of nonlinear-flow-induced fragmentation,
in which case $\vmax = 3.2\,\cs$ and $6.0\,\cs$, respectively.
The bottom panels show the mass-to-flux ratio at the end of
simulations of model 9 and 16. The subcritical model 9 has only
isolated pockets of supercritical cores, as well as emerging cores
which still have subcritical but enhanced mass-to-flux ratio. The
image is similar to the corresponding image when starting with
small-amplitude perturbations (Fig. 9 of BCW), but the
cores are not circular in shape. The fragmentation scale also
seems related to that of the small-amplitude perturbation model, although
many fragments are just beginning to emerge and may 
take a much longer time to develop fully. 
The corresponding image for model 16 shows that the initially
critical ($\mui=1.0$) state leads to many regions of supercritical
mass-to-flux ratio. The emerging fragment pattern 
has less resemblance to the corresponding case that starts from small amplitude
perturbations (Fig. 9 of BCW) than for $\mui=0.5$. Nevertheless,
a fragmentation scale is more apparent than for the case of
nonlinear-flow-induced fragmentation shown in Fig.~\ref{mtoflx}.
The relatively low amount of 
remaining turbulence at the end of both simulations means that new cores will develop
on the non-accelerated ambipolar diffusion time scale, leading to a 
significant age spread of star formation. In the nonlinear
initial condition models, with $k^{-4}$ or $k^0$ spectrum, we may 
identify the initial cores with an early phase of star formation that is
induced by turbulence, and the emerging cores with a later phase
of star formation that can grow from lingering small-amplitude perturbations.
In this way, our model could be in qualitative agreement with the 
empirical scenario advocated by \citet{pal00}. 



\subsection{Time Evolution}

\bfig
\centering
\epsfig{file=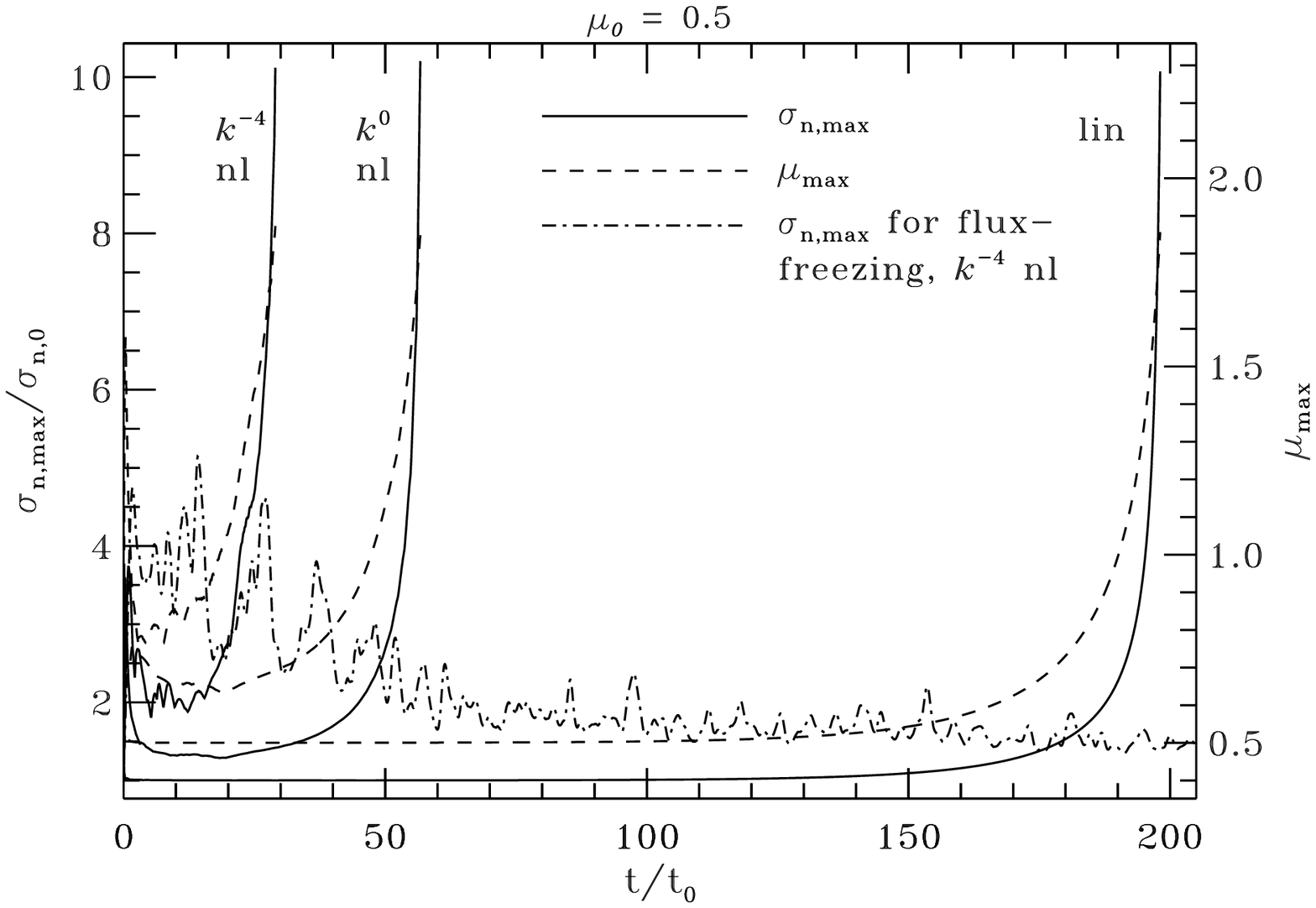,width=1.0\linewidth,clip=} 
\caption{
Time evolution of maximum values of surface density
and mass-to-flux ratio for various models with $\mui=0.5$.
The solid lines show the
evolution of the maximum value of surface density in the simulation,
$\signmax/\signi$, versus time $t/t_0$.
The dashed lines
show the evolution of the maximum mass-to-flux ratio in the simulation,
$\mu_{\rm max}$.
This is shown for models
4, and 9, which have 
$\mui= 0.5$ and same values of $\tniitil$ and $\Pexttil$, 
but different power spectra of turbulent initial perturbations, $k^{-4}$
and $k^0$ respectively. 
Two other models are also shown for comparison. One has the same parameters but
linear initial perturbations, corresponding to model 1 of BCW.
Furthermore, the dash-dotted line shows the evolution (up to time
$t \approx 200\,t_0$ only)
of $\signmax/\signi$ for the flux-frozen ($\tniitil=0$) model 1, which never undergoes
runaway collapse.
}
\label{timevol1}
\efig

\bfig
\epsfig{file=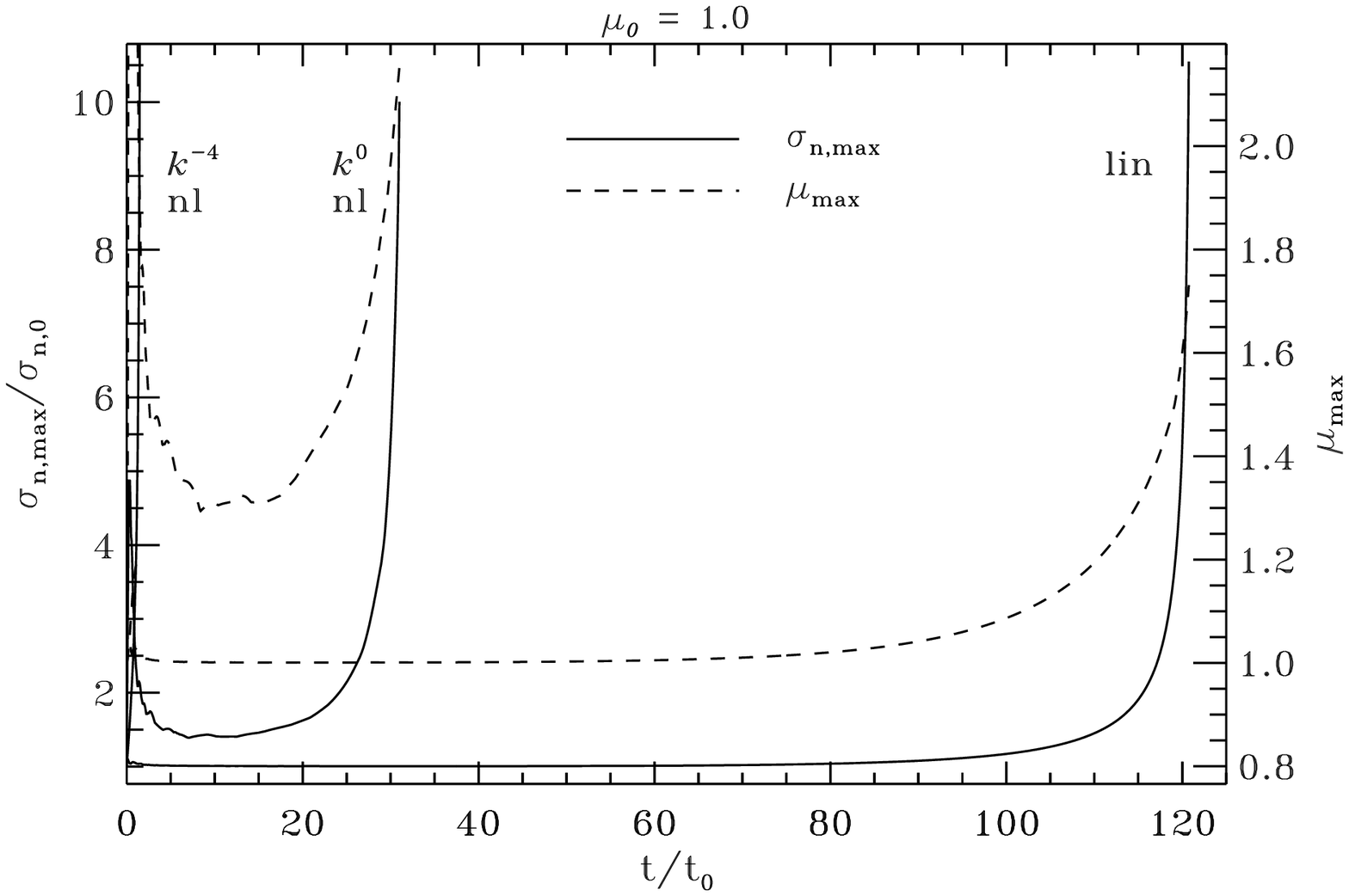,width=1.0\linewidth,clip=} 
\caption{
Time evolution of maximum values of surface density
and mass-to-flux ratio for various models with $\mui=1.0$.
The solid and dashed lines have the same meaning as in Fig.~\ref{timevol1}.
Shown are results from models 13 and 16, which have differing
power spectra of turbulent initial perturbations. Also shown is 
a model with the same parameters but linear initial perturbations,
corresponding to model 3 of BCW. 
}
\label{timevol2}
\efig

\bfig
\epsfig{file=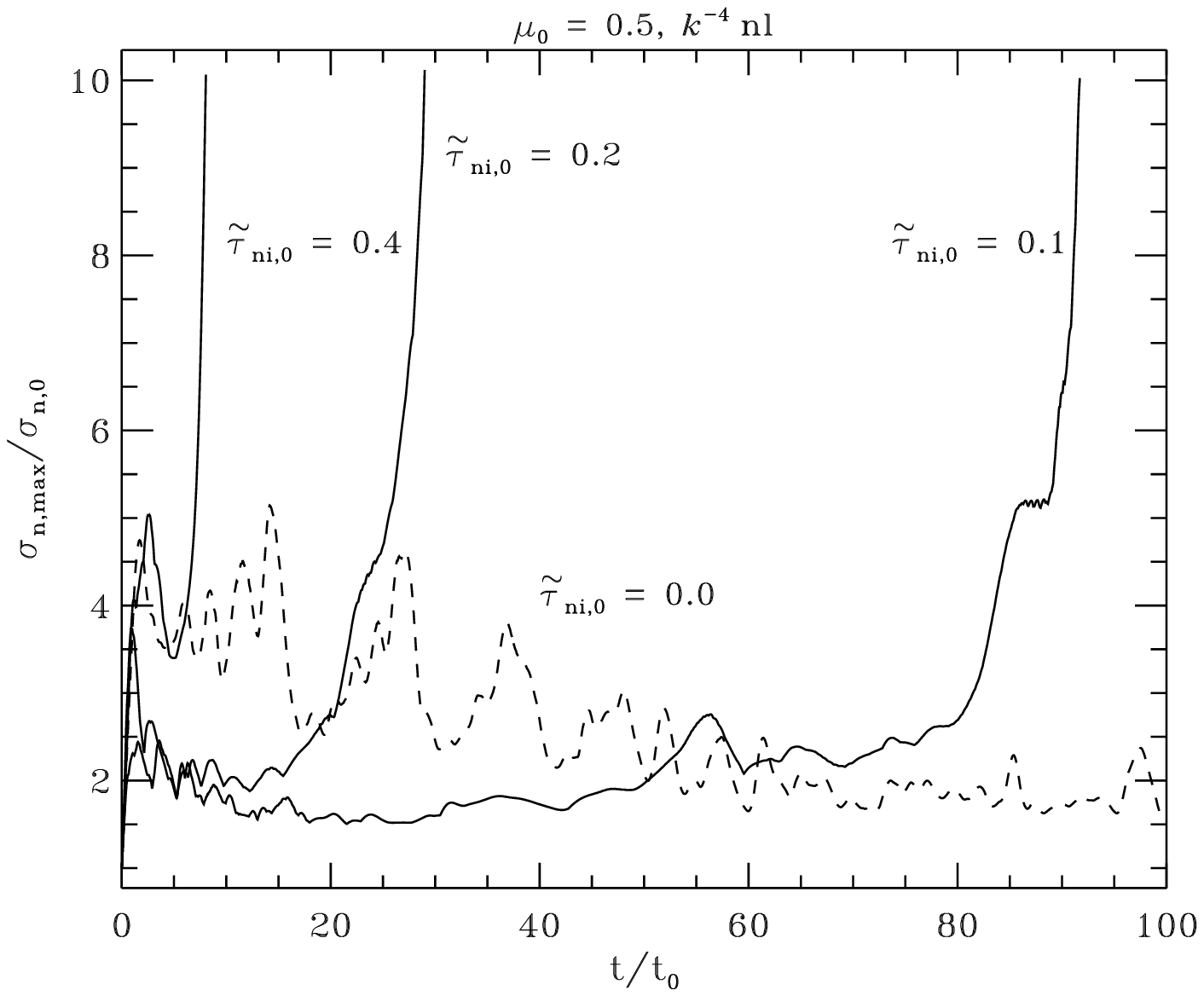,width=1.0\linewidth,clip=} 
\caption{
Effect of differing levels of magnetic coupling on time evolution of
the maximum value of surface density. The solid lines show results for
models 4, 10, and 11. All models have
$\mui=0.5$, $\Pexttil=0.1$, and turbulent initial perturbations with
power spectrum $\propto k^{-4}$. The dashed line shows the evolution of
the flux-frozen model 1 ($\tniitil=0$), which does not undergo runaway
collapse.
}
\label{timevol3}
\efig

Fig.~\ref{timevol1} shows the time evolution of
maximum column density for four models and of the maximum 
mass-to-flux ratio for three models, all having $\mui=0.5$. 
It clearly shows that for subcritical
clouds: (1) fragmentation by runaway collapse does not occur under
conditions of flux-freezing (dash-dotted line; the simulation actually
runs past $t=5000\,t_0$ without runaway collapse); (2) small-amplitude
(linear) perturbations result in the classical quasistatic evolution 
requiring a time $\trun \approx 200\,t_0$; (3) flat spectrum ($k^0$) 
nonlinear perturbations that result in a collapse time that is shorter
by a factor $\approx 4$; (4) power-law ($k^{-4}$) spectrum 
nonlinear perturbations that result in a rapid collapse
that is shorter than the linear case by a factor $\approx 7$.
Of course, the
exact values of $\trun$ will depend on $v_a$ and other parameters
such as $\tniitil$ and $\Pexttil$. For the fourth case above, it may depend
on the box size $L$ as well, since that sets the scale of the largest
mode in the simulation. 
This figure corresponds to Fig. 1 of \citet{kud08}, which 
shows results for some three-dimensional models. In their figure,
the volume density $\rhon$ and plasma $\beta$ (counterparts to 
$\sign$ and $\mu$ in our thin-sheet model) undergo some variations 
due to vertical oscillations of the cloud, before eventually increasing rapidly.
Our model follows the integrated quantities through the layer and
therefore does not include the effect of vertical oscillations. 
Nevertheless, the timescale of
evolution and eventual runaway collapse are in good agreement 
where comparisons can be made. Our thin-sheet model allows a 
broader parameter study than 
currently possible using three-dimensional simulations. 

Fig.~\ref{timevol2} shows the time evolution of maximum column density and
maximum mass-to-flux ratio for three models with $\mui=1.0$. 
The model with small amplitude (linear) initial perturbations 
corresponds to model 3 of BCW. The other ones are 
model 13 and model 16 of this paper. 
The same three qualitatively distinct evolutionary modes occur
as for the case of $\mui=0.5$. A notable difference is that 
collapse occurs immediately during the first compression
in the case of nonlinear-flow-induced fragmentation, at 
$\trun=1.8\,t_0$. There is no rebound from the first compression
as occurs when $\mui=0.5$.

Fig.~\ref{timevol3} illustrates the effect of varying levels of ionization
on the fate of nonlinear-flow-induced fragmentation. This is 
represented by differing values of $\tniitil$, with $\tniitil=0$
corresponding to flux-freezing and differing values of 
$\tniitil$ corresponding to different initial ionization fractions
$x_{\rm i,0} \propto \tniitil^{-1}$
(see Appendix).
The standard model with $\tniitil=0.2$ corresponds to 
a canonical ionization fraction $x_{\rm i,0} \simeq 10^{-7}(\nni/10^4\cmc)^{-1/2}$ 
\citep{tie05}.
Our results show that $\trun$ may indeed vary significantly 
due to variations in $x_{\rm i,0}$, easily spanning the range of
$10^6$ yr to $10^7$ yr for typical values of units and input parameters.
Clearly, a definitive understanding of the influence of magnetic 
fields awaits further insight into ionization levels in molecular clouds.


\subsection{Core Properties}

\bfig
\epsfig{file=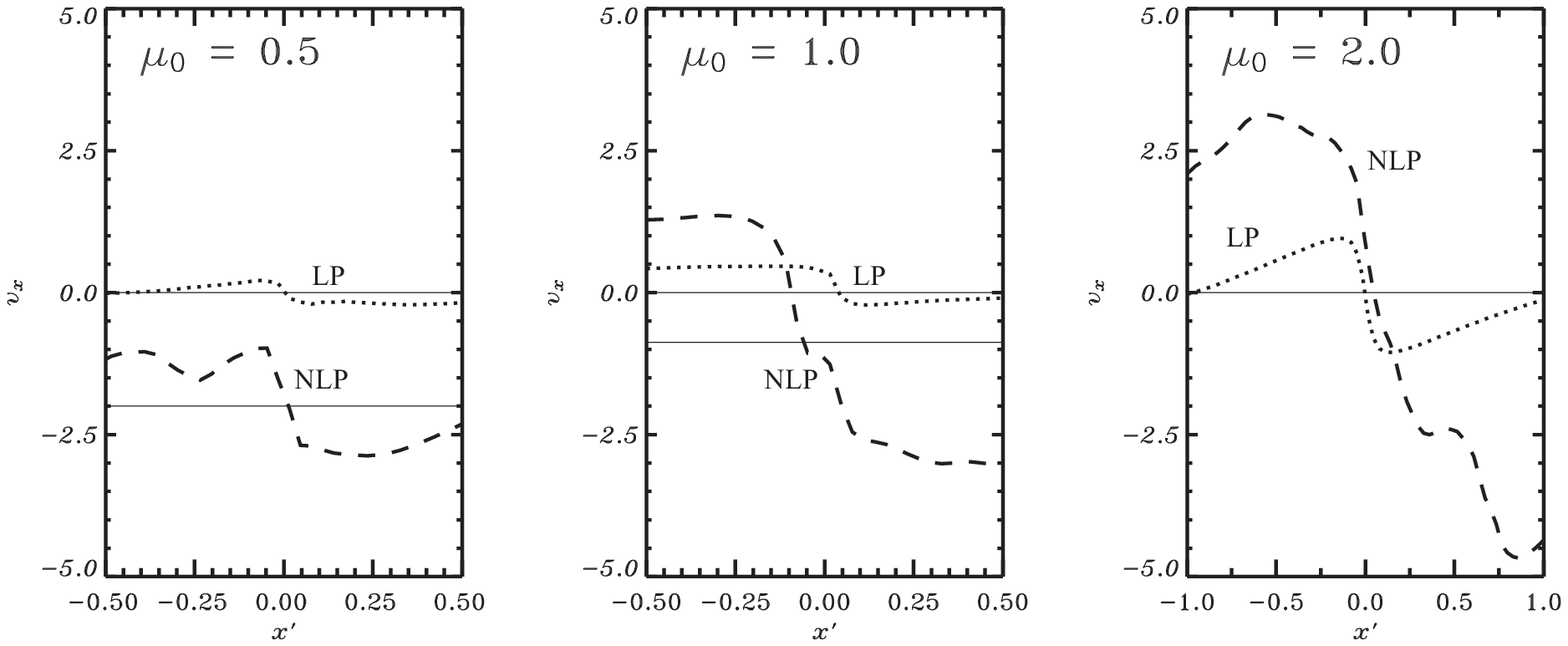,width=1.0\linewidth,clip=}
\caption{
Velocity component of neutrals $v_x$, normalized to $\cs$, 
along a line parallel to the $x$-axis
that passes through the center of a core, for models with
various values of $\mui$ and differing initial perturbations.
The horizontal coordinate $x' = (x - x_{\rm c})/L_0$, where $x_{\rm c}$
is the location of the core center in each case.
The dashed lines show the profile of $v_x$ for cores generated in models
4, 13, and 17, from left to right panels. They are characterized by
initial nonlinear perturbations (NLP). The dotted lines show
for comparison the profiles through cores in models 1, 3, and
5 of BCW, i.e. models with the same parameters
but initial linear perturbations (LP). 
The horizontal solid lines denote the systematic speeds of the
cores in the $x-$direction. Note the largest systematic speed
in the model with $\mui=0.5$ and initial NLP.
}
\label{velcuts}
\efig

Two important observed properties of dense cores are the
kinematics of infall motions, and the
distribution of core masses. The latter suffers from some
ambiguity due to the different possible definitions of 
a ``core'', or more specifically, how to define a core
``boundary''. Here we describe the most basic features of
our simulated cores.

Fig.~\ref{velcuts} shows the velocity profiles (using dashed lines) in the 
vicinity of cores that are obtained in simulations
of models 4, 13, and 17, with $\mui=0.5,1.0$, and 2.0,
respectively. They are measured along 
a line in the $x$-direction that passes through 
the column density peak. Also shown (in dotted lines) for
each value of $\mui$ is a profile of $x$-velocity through a core
that is formed in a simulation with small-amplitude (linear)
perturbations (BCW), but otherwise the same parameters as the other
model shown in the same panel.
The horizontal solid lines in each panel pass through the 
``midplane'' of the velocity profiles, and allow one to read off the
systematic $x$-velocity of the core. 
For the cases of initial linear perturbations, the increasing 
sequence of $\mui$ leads to ever increasing maximum infall speeds,
from about half the sound speed up to mildly supersonic values.
There is also evidence of infall speeds increasing
towards the core centers, due to gravitational acceleration.
For the cases of initial nonlinear perturbations (these models all
have a $k^{-4}$ spectrum), the sequence of increasing $\mui$ leads to 
greater relative infall speeds onto the cores. These motions are
supersonic in all cases, and constitute an important observationally-testable
consequence of nonlinear-flow-induced fragmentation. 
The models with greater $\mui$ have greater infall speeds because they
undergo collapse during the first compression, with most of
the initial input turbulent energy still intact, i.e. there has not
been much time for turbulent decay. Also note that there is no 
evidence for an accelerating flow in these cases, which would be
a signature of gravitationally-driven motions. 
Thus, these models demonstrate flow-driven core formation, rather than
gravitationally-driven core formation.
For the model with
$\mui=2.0$ there is essentially no systematic core speed, since the 
collapse occurs very quickly at the intersection of two
colliding flows. At the other limit of a significantly subcritical
cloud ($\mui=0.5$), the initial compression is followed by a rebound
due to the strong magnetic restoring forces. The core forms later
within the region of high density that is undergoing oscillatory motions.
The systematic speed of the core relative to the simulation box 
is supersonic (about twice the sound speed in this model), although
the relative speed of infall onto the core is subsonic or transonic.
The large systematic core speeds for subcritical clouds constitute another 
observationally-testable
consequence of nonlinear-flow-induced fragmentation.
The case of $\mui=1.0$ is intermediate in features between the
$\mui=0.5$ and $\mui=2.0$ models, but is actually closer to the $\mui=2.0$
model since the collapse occurs very quickly, with $\trun=1.8\,t_0$.
This is very close to the value $\trun=1.3\,t_0$ for the $\mui=2.0$ model.



\bfig
\epsfig{file=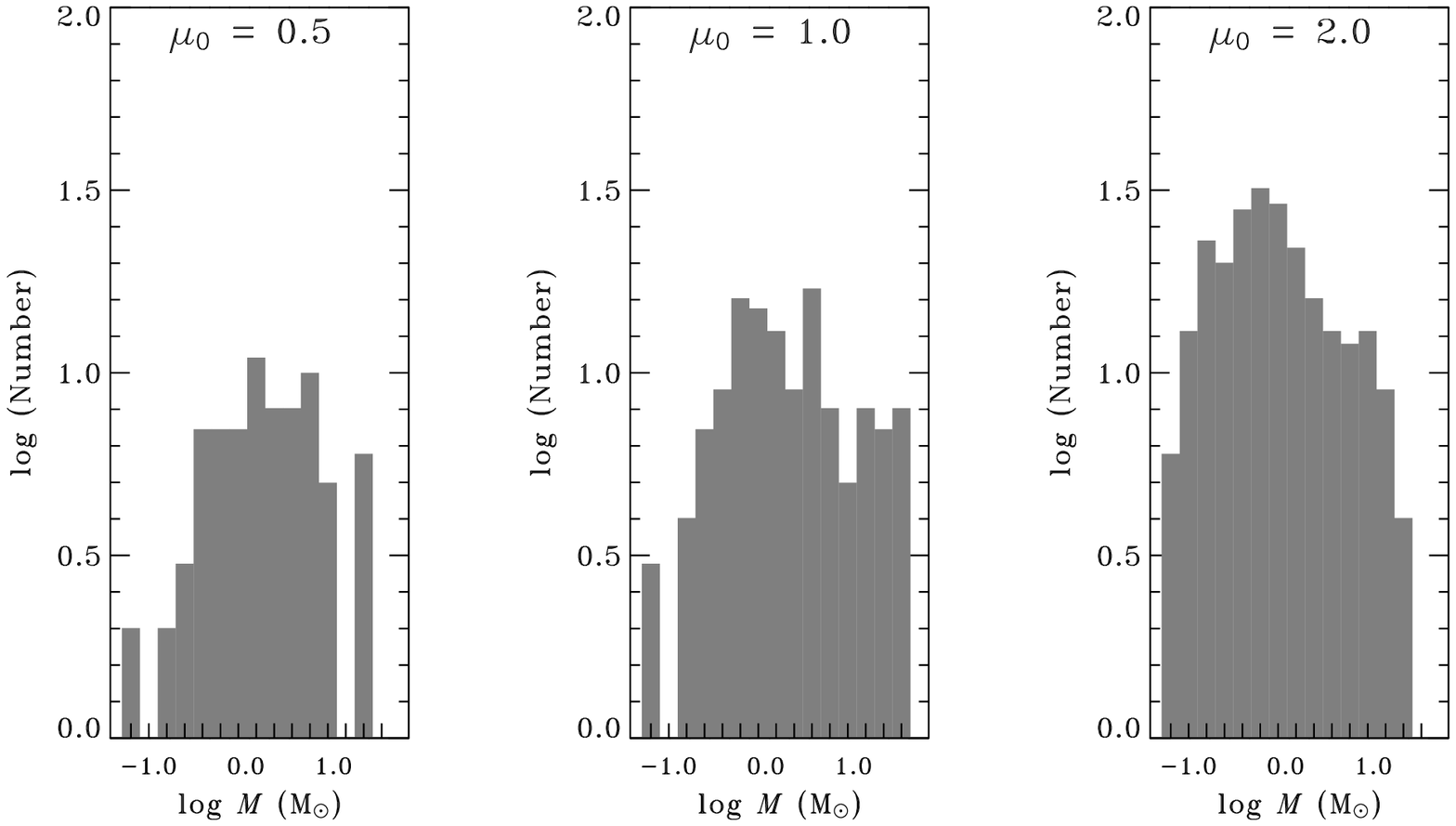,width=1.0\linewidth,clip=}
\caption{
Histograms of masses contained within regions with $\sign/\signi \geq 2
$,
measured at the end of simulations with parameters of
models 4, 13, and 17. Specifically, they are distinguished by
values of $\mui =  0.5,1.0,2.0$ as labeled.
Each figure is the result of a compilation of results of many
simulations. The bin width is 0.1.
}
\label{masshist}
\efig

Fig.~\ref{masshist} shows the histograms of core masses, defined as masses
enclosed within regions that have $\sign/\signi \geq 2$ surrounding a 
column density peak. These are measured at the end of each simulation,
for 15 separate realizations of model 4, and 25 each of models 13 and 17. 
Simulations of model 4 ($\mui=0.5$) and model 13 ($\mui=1.0$)
produce an average of five identifiable cores per simulation, 
while model 17 ($\mui=2.0$) produces an average of ten cores per simulation.
For details
about our thresholding technique used to obtain core masses, see BCW.
The histograms reveal that for any fixed value of $\mui$, the distribution
of core masses is much broader than the corresponding histogram of masses
for fixed $\mui$ and initial small-amplitude (linear) perturbations.
See Fig.~8 of BCW for the latter, which show a very sharp descent
beyond the preferred mass scale. There are many more high-mass cores that
are formed in these models. However, an examination of Figs. \ref{densimgs}
and \ref{movieimgs} reveals that many of the cores have very elongated
and irregular shapes, so that they may yet break up into multiple fragments.
In BCW, we proposed that a broad CMF may be caused by a distribution of 
initial mass-to-flux ratio values
within a cloud. That remains an alternative scenario to that of 
``turbulent fragmentation'' explored here and in several previous
publications \citep{pad97,kle01,gam03,til07}.

\subsection{Turbulent Dissipation}
\label{s:tdiss}

\bfig
\epsfig{file=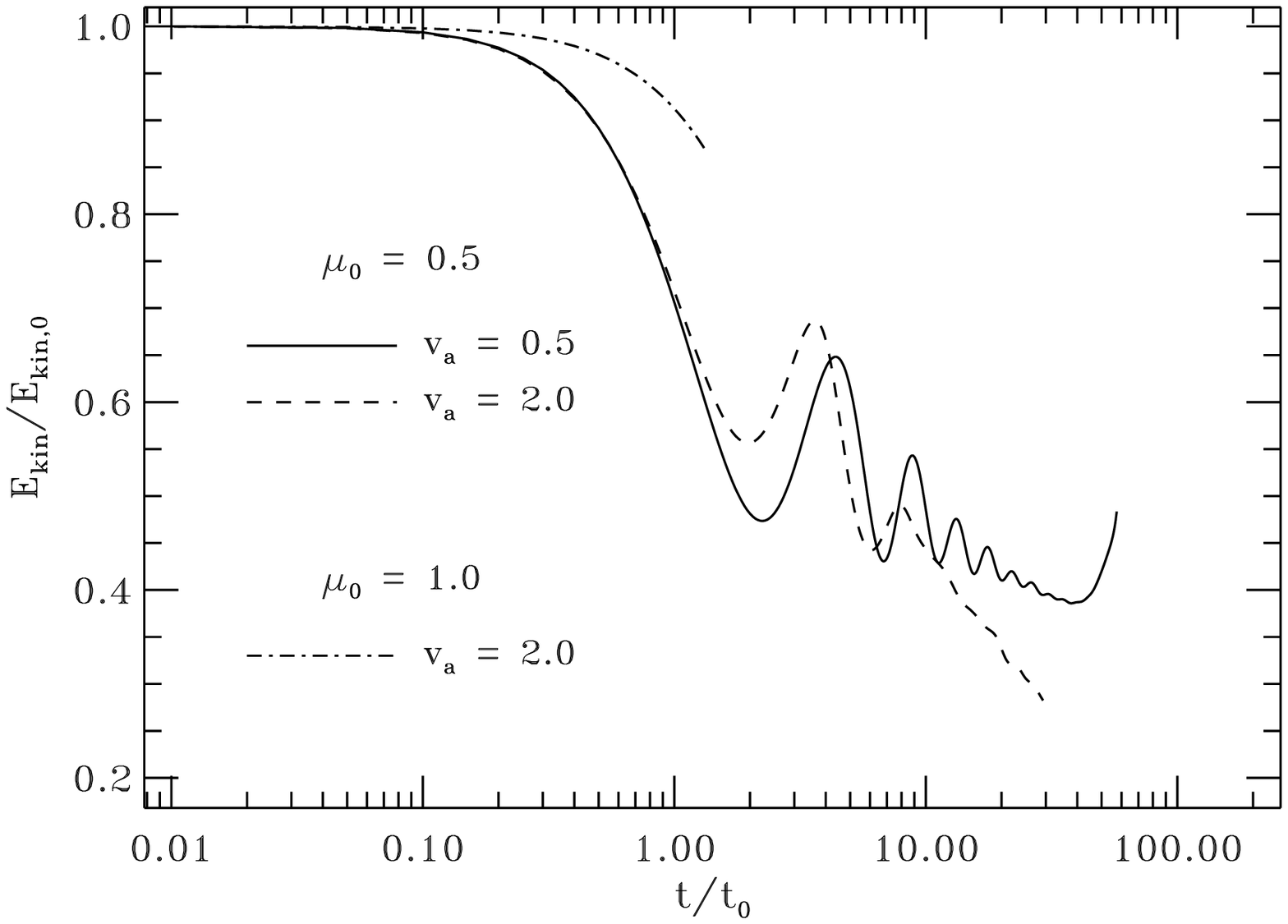,width=1.0\linewidth,clip=}
\caption{
Kinetic energy versus time for models with various values
of $\mui$ and/or $v_a$ (normalized to $\cs$).
Each of these models is run with $N=256$.
}
\label{ekin1}
\efig

The rate of dissipation of turbulent energy has been studied
extensively in a series of three-dimensional simulations
\citep[e.g.][]{sto98,mac98,mac99,ost01}. See also the reviews by 
\citet{mac04}, \citet{elm04}, and \citet{mck07}.
In this Section, our goal is to briefly present some 
information about the turbulent decay in our simulations. 
These may be of interest because our simulations differ in their
use of the thin-sheet approximation and the use of high-order adaptive
time-stepping that is part of our implementation of the method of lines 
technique (see BCW).
Nevertheless, we do also obtain relatively rapid turbulent
dissipation in most models, as presented in the various figures in this Section.

We present results for the decay of kinetic energy in our simulation
box. However, the total energy in the simulation box
is not conserved, due to radiative losses implied by our isothermal
assumption, and also due to work done by the external pressure and 
magnetic forces associated 
with the field components $B_x$ and $B_y$ at the
cloud's top and bottom surfaces. Nevertheless, the amount of turbulent
kinetic energy present in a cloud has important observational
implications, so we illustrate its evolution here in several figures.
We also use the kinetic energy evolution as a means of exploring the
effect of numerical resolution in our simulations.

Fig.~\ref{ekin1} shows the evolution of kinetic energy $\ekin$
(defined as the sum of $\frac{1}{2} \sign(v_x^2+v_y^2)$ over all
cells of a simulation), normalized to
initial values $\ekini$, for models 4, 6, and 13. 
Note that the values of $\ekini$ differ from model to model. 
The model with $\mui=1.0$ does not have much chance to lose kinetic energy
because collapse occurs right away, during the first
turbulent compression. For models with $\mui=0.5$, there is a rebound from
the initial compression, and this is indicated by the 
oscillations of $\ekin$. Furthermore, there is an overall systematic
decay of $\ekin$ so that it is significantly reduced in one sound crossing
time of the initial half-thickness of the cloud,
$t_{\rm c} = Z_0/\cs \simeq 2 L_0/\cs = 2\,t_0$, where we have used
Eq. (30) of BCW to relate $Z_0$ to $L_0$. The decaying oscillations
of $\ekin$ are consistent with the qualitative picture obtained from the 
animation of model 4 that accompanies Fig.~\ref{movieimgs}. Some of our
realizations do show an increase in $\ekin$ during the last stage of evolution,
due to the conversion of gravitational energy into kinetic energy
of systematic infall onto one or more cores.

Fig.~\ref{ekin2} shows the effect of resolution on the decay of kinetic energy.
Our standard simulations have $N=128$, and numerical experiments with
$N=256$ and $N=512$ demonstrate that while some more kinetic energy is 
retained in those cases, the overall pattern of decay and oscillations
of $\ekin$ is maintained. 
We note that each simulation has a unique random but statistically
equivalent initial state.

Fig.~\ref{ekin3} reveals the additional effect on the kinetic energy evolution 
of two interesting limits. 
In one case, we perform the numerical experiment of starting with 
a divergence-free (non-compressive) initial velocity field, 
even though turbulence in the interstellar medium is thought to 
be highly compressive \citep{mck07}.
For this case, a plot of $\ekin$ for the evolution of
model 8 (dashed line) reveals that the
kinetic energy still decays, but that the cloud does not undergo 
large-scale oscillations during the process. These oscillations occur in the
compressible case due to the restoring force of the magnetic field when 
compressed into filamentary structures. This does not occur in the 
incompressible model in a globally coherent manner, although locally 
compressive motions are generated during the evolution and turbulent decay
does occur rapidly, as in the other models. 
The dash-dotted line reveals the interesting evolution in the case of 
flux-freezing (model 1; $\tniitil=0$). Here the initially compressive 
velocity field leads to an initial rapid decay of turbulence through
shocks, but large-scale oscillatory modes 
remain in the simulation box for indefinite periods of time.
These modes have a root mean squared velocity amplitude $\approx 2\,\cs$
and contain roughly half the initial input energy.
Why do these modes not decay away? The lack of ambipolar diffusion
means there is no dissipation of modes in which the restoring force is
due to the magnetic field. The $k^{-4}$ spectrum means that the largest
modes dominate, and these also suffer negligible numerical dissipation
in our scheme. The restoring force that drives the waves is provided largely
by the magnetic tension associated with the magnetic field external to the
sheet. While the case of a thin sheet may not be generalizable to
three dimensional molecular clouds, we feel this result is an 
important pointer to processes that may in fact be occurring in real 
clouds. That is, the external magnetic field, anchored in the Galactic
interstellar medium, may allow the outer parts of clouds 
(effectively flux-frozen due to UV ionization - see \citealt{cio95})
to maintain long-lived oscillations
that are then identified observationally as ``turbulence''. 
This idea has long been advocated by \citet{mou75,mou87}.
We note that this result could not be obtained in periodic box simulations
that contain no effect of an external medium, and leave a 
more thorough assessment of this effect to a forthcoming paper.


\bfig
\epsfig{file=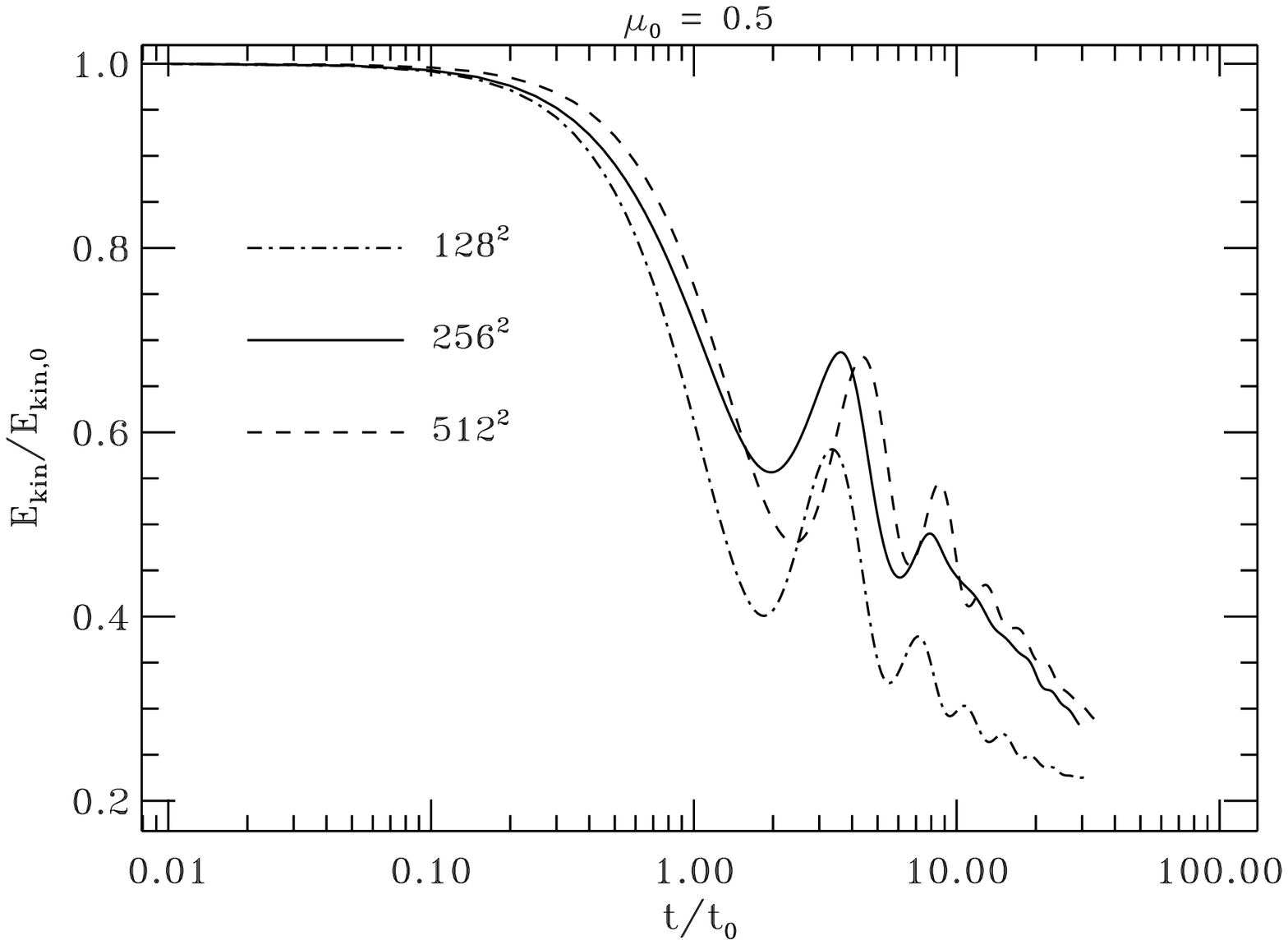,width=1.0\linewidth,clip=}
\caption{
Kinetic energy versus time for model 4 parameters but varying resolution.
}
\label{ekin2}
\efig

\bfig
\epsfig{file=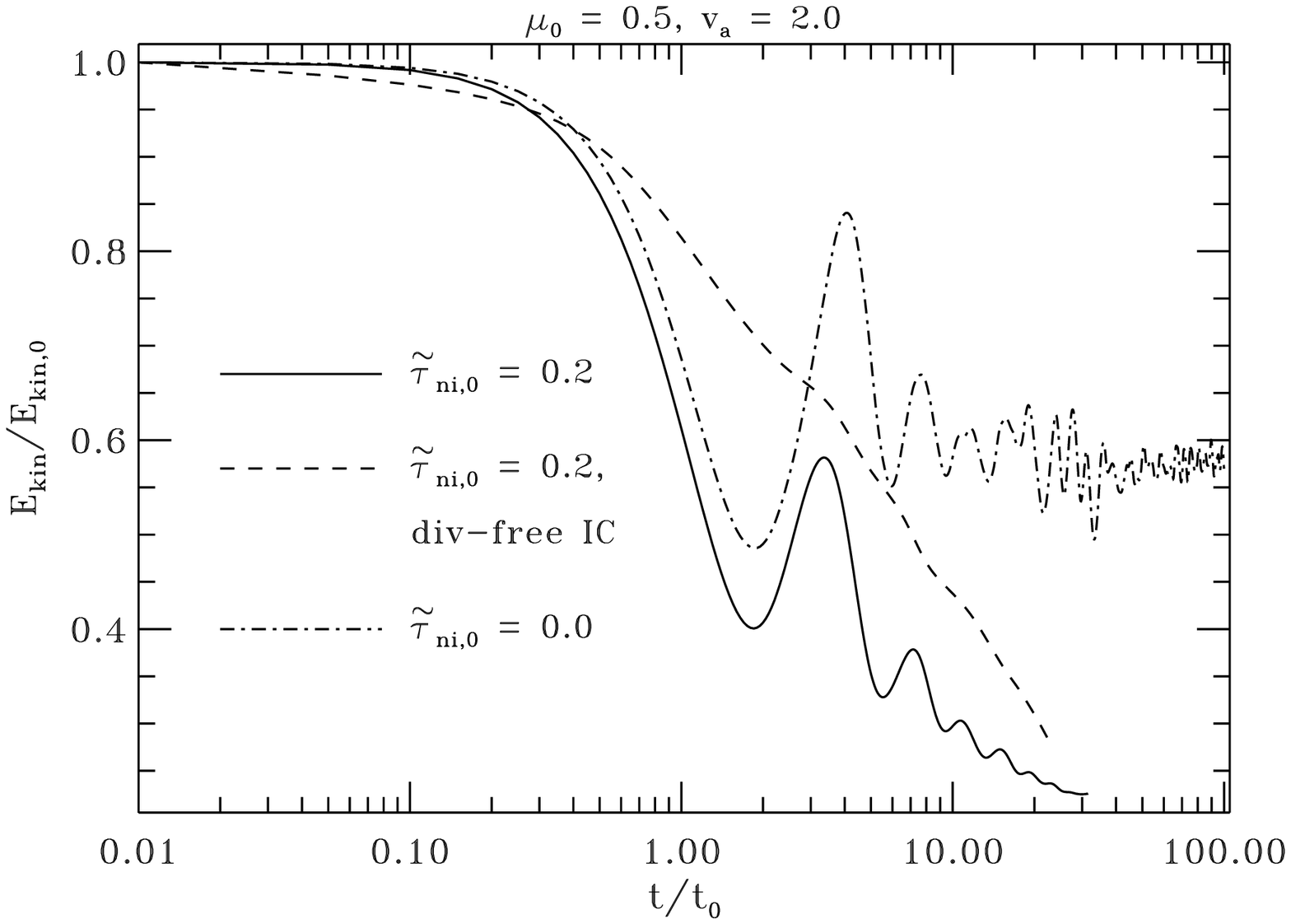,width=1.0\linewidth,clip=}
\caption{
Kinetic energy versus time for models 1 (dash-dotted line), 4 (solid line),
and 8 (dashed line).
Each of these models has $\mui=0.5$ and $v_a = 2\,\cs$ and
is run with $N=128$. They differ in their values of 
$\tniitil$ and in whether or not the initial velocity
field is divergence-free.
}
\label{ekin3}
\efig

\section{Discussion}

We have performed a parameter study of fragmentation of a dense sheet
aided by the presence of initial nonlinear velocity perturbations.
In most models, the power spectrum of fluctuations is $\propto k^{-4}$,
so that the initial conditions impose primarily a large-scale flow
to the system. We have also studied the case of nonlinear perturbations
with power spectrum $\propto k^0$, in which the small-scale fluctuations
play a bigger role. Of the two modes, the latter is more similar to
gravitational fragmentation arising from small-amplitude perturbations,
as studied extensively in our previous paper (BCW). The main
difference is an accelerated time scale for core formation. This is
particularly apparent for the cases with subcritical initial 
mass-to-flux ratio, in which case the nonlinear fluctuations enhance
ambipolar diffusion \citep[see][]{fat02,zwe02}. For the 
case of nonlinear-flow-induced fragmentation, originally studied
by \citet{li04} and \citet{nak05}, we find that the induced structures
are highly filamentary and go into direct collapse for supercritical
clouds. For critical, and more so for subcritical clouds, the 
initial compression may be followed by a rebound and oscillations
which eventually lead to runaway collapse in dense pockets where
enhanced ambipolar diffusion has created supercritical conditions.
What determines the outcome? For any given field strength, there is
a threshold initial velocity amplitude $v_a$ above which prompt collapse
will take place. 
An examination of model outcomes in Table 1 reveals that, for a fixed
standard initial ionization fraction defined by $\tniitil=0.2$, 
prompt collapse takes place when $v_a > \vms$
(see models 2, 13, and 17). Indeed, the model 3, which has 
$v_a \approx \vms$, 
is actually prone to go into prompt collapse ($\trun \approx t_0$) 
in some realizations, 
but undergoes several oscillations before runaway collapse in
most cases (with representative value $\trun=30\,t_0$). 
We can say that significantly super-\Alfvc perturbations are associated
with prompt collapse, for both subcritical and supercritical model clouds.
This criterion does not apply to models with initial power
spectrum $\propto k^0$,
since the kinetic energy does not get channeled toward a large-scale
compression wave. It also does not apply to model 7 ($\Pexttil = 10$),
since its low value of $\vms$ is very specific to the 
external-pressure-dominated initial state, but not representative of 
the signal speed in the high-density regions that are subsequently generated.


The highly filamentary structure of clouds in which prompt collapse
takes place is a source of concern when comparing with maps of observed
molecular clouds. This was noted by \citet{li04}, who commented that
clouds with weak magnetic field and supersonic turbulence as modeled
(i.e. $k^{-4}$ power spectrum, having most power on the largest scales) 
would appear too filamentary in comparison with observations.  
Our study extends this concern also to models with strong magnetic field
if the turbulence is highly super-\Alfv, since prompt collapse occurs
in highly compressed filaments without a chance for them to rebound.
Since the weak magnetic field cases also have by design a velocity 
amplitude that is super-\Alfv, we can say that super-\Alfvc turbulence
in all cases may have problems with excessive filamentarity. There is another problem
with large amounts of turbulent forcing; the relative infall motions onto
the cores are highly supersonic (see Fig.~\ref{velcuts}), and at odds with
observed core infall motions \citep{taf98,wil99,lee01,cas02}, which are 
subsonic or at best transonic. 
Of course, both of these problems are set up artificially in our
simulations through nonlinear forcing associated with the initial conditions.
In other simulations of driven super-\Alfvc turbulence \citep{pad02}, such
forcing continues at all times and the above features are always present.

If the highly turbulent and/or super-\Alfvc models pose difficulties for
dense core formation, then 
how does one account for the highly supersonic motions observed in molecular clouds
\citep[e.g.][]{sol87}? 
The answer is likely that they exist in the lowest density envelopes of the 
molecular clouds and therefore should not be input into local models of dense
subregions, as we do in some cases here. Our super-\Alfvc models in a periodic
simulation box demonstrate a limiting case, and help establish that such models
cannot be applied directly to explain observed star-forming regions.
In a global scenario, the dense regions
that form cores will be less turbulent than the larger low-density envelopes.
The low-density regions can support highly turbulent motions while denser regions
have lower velocity dispersion, as demonstrated in 1.5 dimensional global
models of molecular cloud turbulence \citep{kud03,kud06}.

Our Fig.~\ref{masshist} shows that the core mass distribution is relatively 
broad for any given value of $\mui$ for nonlinear-flow-induced
fragmentation  ($k^{-4}$ spectrum of velocity fluctuations). 
This repeats the qualitative findings of many earlier studies 
in three-dimensions \citep{pad97,kle01,gam03,til07}.
This scenario of turbulent fragmentation is a plausible mechanism to 
generate broad CMFs of the type observed. It remains an open question whether
this kind of CMF is related to the IMF since the cores are often highly irregular
in shape, and it is not clear that they will collapse monolithically.
Alternative methods to generate broad IMFs or CMFs are the global effect of 
competitive accretion \citep{bon03,bat03}, a temporal spread of core
accretion lifetimes \citep{mye00,basj04}, or a distribution of initial
mass-to-flux ratios in a cloud (BCW).  
Future work by the astrophysical community may clarify the relative 
roles of these processes.




\section{Summary}

We have studied the 
effect of initial nonlinear velocity perturbations on the 
formation of dense cores in isothermal sheet-like layers that 
may be embedded within larger molecular cloud envelopes. 
Our simulation box is periodic in the lateral ($x,y$) directions and 
typically spans four nonmagnetic (Jeans) fragmentation scales in each of these
directions. The initial input turbulent energy is allowed to decay freely.
The simulations reveal a wide range of outcomes.
We emphasize the following main results of the paper:

\begin{enumerate}

\item{\it Time Evolution to Runaway.}
Subcritical model clouds can undergo accelerated ambipolar 
diffusion in two different ways. For nonlinear
initial velocity perturbations in which small-scale modes contain
a large portion of the energy, the onset of runaway collapse 
occurs sooner by a factor $\approx 4$ in our typical models.
For nonlinear perturbations with most energy on the largest scales
(hereafter, nonlinear flows),
the runaway collapse can be sped up
by a greater factor, $\approx 7$ for our typical models.
Supercritical clouds undergo prompt collapse
whenever nonlinear flows are present.
Subcritical model clouds may also be pushed into prompt 
collapse by nonlinear flows that are significantly super-\Alfv. 

\item{\it Morphology of Clouds.}
Supercritical model clouds whose evolution is initiated by nonlinear
flows have a highly filamentary structure. Subcritical clouds with 
initial nonlinear but trans-\Alfvc or sub-\Alfvc flows have a markedly
less filamentary structure. In these cases, magnetic fields
cause a rebound from the initial compression, and several oscillations
occur before the runaway collapse of the first cores.
Subcritical clouds with initially super-\Alfvc nonlinear flows promptly develop
highly filamentary structure with embedded collapsing cores.

\item{\it Velocity Profiles.}
Supercritical and transcritical model clouds which are driven into prompt collapse
have highly supersonic infall speeds at the core boundaries,
while subcritical model clouds typically have transonic or subsonic 
infall speeds (relative to the velocity centroid) onto cores.
In the subcritical cases, the cores can have larger systematic motions than in
supercritical models, because the cores form within regions undergoing oscillatory
motions.
We believe that the large infall motions in the models with super-\Alfvc 
nonlinear flows may disqualify them as viable models for core formation,
given current observational results.

\item{\it Core Mass Distributions.}
Core formation initiated by nonlinear flows leads to broader core mass functions
than found in earlier studies of fragmentation initiated by small-amplitude
perturbations. This applies to models of any fixed initial mass-to-flux
ratio $\mui$.
However, the ultimate relation of such a core mass function to the 
stellar initial mass function is not settled due to the irregular shape of
the fragments created by nonlinear flows.
These fragments may in turn break up into multiple objects at a later stage.

\item{\it Turbulent Decay.}
Supersonic initial velocity perturbations lead to an initially rapid
decay of kinetic energy in all models, on a time scale similar to the sound crossing
time across the half-thickness of the sheet. This rapid decay of turbulence
is in agreement with a wide variety of previous results in the literature.
However, subcritical model clouds can undergo oscillations that 
reduce the decay rate of kinetic energy at later times. Furthermore, in the
limit of excellent neutral-ion coupling (flux-freezing), as may be present
in UV-ionized molecular cloud envelopes, large-scale wave modes may survive
for very long times. 

\end{enumerate}

\section*{Acknowledgements}
We thank the anonymous referee for comments which improved 
the discussion of results.
We also thank Stephanie Keating for creating color images and animations
using the IFRIT package developed by Nick Gnedin.
SB was supported by a grant from the Natural Sciences and Engineering
Research Council (NSERC) of Canada. 
JW was supported by an NSERC Undergraduate Summer Research Award.

\appendix

\section{Typical Values of Units and Other Quantities}

Typical values of our units are
\barray
\label{cs}
\cs & = & 0.188 \, \left(\frac{T}{10 \K}\right)^{1/2} \kms,  \\
\label{t0}
t_0 & = & 3.65 \times 10^4 \left(\frac{T}{10\,\K}\right)^{1/2}
\left(\frac{10^{22}\,\cms}{\Nni}\right) \yr, \\
\label{L0}
L_0 & = & 7.02 \times 10^{-3} \left(\frac{T}{10 \K}\right)
\left(\frac{10^{22}\,\cms}{\Nni}\right) \pc \\ \nonumber
& = & 1.45 \times 10^3  \left(\frac{T}{10 \K}\right)
\left(\frac{10^{22}\,\cms}{\Nni}\right) \AU, \\
\label{M0}
M_0 & = & 9.19 \times 10^{-3} \left(\frac{T}{10\,\K}\right)^2
\left(\frac{10^{22}\,\cms}{\Nni}\right) \Msun, \\
\label{B0}
B_0 & = & 63.1 \left(\frac{\Nni}{10^{22}\,\cms}\right) \muG.
\earray
Here, we have used $\Nni = \signi/\mn$, where $\mn = 2.33\,\mH$ is the mean
molecular mass of a neutral particle for an H$_2$ gas with a 10\%
He abundance by number.
Furthermore, we may calculate the number density of the background state as
\beq
\nni = 2.31 \times 10^5
\left(\frac{10 \K}{T}\right)
\left(\frac{\Nni}{10^{22}\cms}\right)^2
\left(1+\Pexttil\right)\, \cmc.
\eeq
The dimensional background reference magnetic field strength for a given model
is simply
$\Bref = B_0/\mui$.
Finally, the ionization fraction ($=\nion/\nn$) in the cloud
may be expressed as
\beq
\label{ioniz}
\xion = {\cal K} \nn^{-1/2} =  3.45 \times 10^{-8} \left(\frac{0.2}{\tniitil}\right) \left(\frac{10^5 \cmc}{\nn}\right)^{1/2} \left(1+\Pexttil\right)^{-1/2}.  
\eeq

\end{document}